\documentclass[11pt,preprint]{aastex}
\pdfoutput=1
\usepackage{graphicx}
\usepackage{lscape}

\let\oldfootsep=\footnotesep
\newcommand\ltsima{$\; \buildrel <\over\sim \;$}
\newcommand\simlt{\lower.5ex\hbox{\ltsima}}
\newcommand\gtsima{$\; \buildrel >\over\sim \;$}
\newcommand\simgt{\lower.5ex\hbox{\gtsima}}
\newcommand\etal{et~al.}

\newcommand\msun {M_\odot}

\newcommand\pac{Paczy{\'n}ski }

\newcommand{\mathbold}[1]{\mbox{\boldmath $\bf#1$}}
\newcommand\piEbold{{\mathbold \pi_E}}

\shorttitle{}
\shortauthors{Bennett et al}

\begin{document}


\title{The First Planetary Microlensing Event with Two Microlensed Source Stars}


\author{D.P.~Bennett\altaffilmark{1,2,M,P},
A.~Udalski$^{3,O}$,
C.~Han$^{4,\mu}$,
I.A.~Bond\altaffilmark{5,M}, 
J.-P.~Beaulieu$^{6,P}$,
J.~Skowron$^{3,O}$,
B.S.~Gaudi$^{7,\mu}$,
N.~Koshimoto$^{1,8,M}$,
\\
and \\
F.~Abe$^{9}$, 
Y.~Asakura$^{9}$, 
R.K.~Barry$^{1}$, 
A.~Bhattacharya\altaffilmark{1,2},
M.~Donachie$^{10}$,
P.~Evans$^{10}$,
A.~Fukui$^{11}$, 
Y.~Hirao$^{8}$, 
Y.~Itow$^{9}$,  
M.C.A.~Li$^{10}$,
C.H.~Ling$^{5}$, 
K.~Masuda$^{9}$,  
Y.~Matsubara$^{9}$, 
Y.~Muraki$^{9}$, 
M.~Nagakane$^{8}$,
K.~Ohnishi$^{12}$, 
H.~Oyokawa$^{9}$, 
C.~Ranc$^{1}$,
N.J.~Rattenbury$^{10}$, 
M.M.~Rosenthal$^{13}$, 
To.~Saito$^{14}$,
A.~Sharan$^{10}$,
D.J.~Sullivan$^{15}$, 
T.~Sumi$^{8}$,
D.~Suzuki$^{16}$,
P.J.~Tristram$^{17}$,
A. Yonehara$^{9}$,
 \\ (The MOA Collaboration)\\
M.K.~Szyma{\'n}ski$^3$,
R.~Poleski$^{3,7}$,
I.~Soszy{\'n}ski$^3$,
K.~Ulaczyk$^3$,
{\L}.~Wyrzykowski$^3$, \\ (The OGLE Collaboration)\\
D.~DePoy$^{18}$,
A.~Gould\altaffilmark{7,19,20},
R.~W.~Pogge$^{7}$
J.C.~Yee,$^{21}$ \\ (The $\mu$FUN Collaboration)\\
M.D.~Albrow$^{22}$,
E.~Bachelet$^{23}$,
V.~Batista$^{6}$,
R.~Bowens-Rubin$^{24}$,
S.~Brillant$^{25}$,
J.A.R.~Caldwell$^{26}$,
A.~Cole$^{27}$,
C.~Coutures$^{6}$,
S.~Dieters$^{27}$,
D.~Dominis~Prester$^{28}$,
J.~Donatowicz$^{29}$,
P.~Fouqu\'e$^{30,31}$,
K.~Horne$^{32}$,
M.~Hundertmark$^{32,33}$,
N.~Kains$^{34}$,
S.R.~Kane$^{35}$,
J.-B.~Marquette$^{6}$,
J.~Menzies$^{36}$,
K.R.~Pollard$^{22}$,
C.~Ranc$^{1}$,
K.C.~Sahu$^{34}$,
J.~Wambsganss$^{37}$,
A.~Williams$^{38,39}$, and
M.~Zub$^{37}$
\\ (The PLANET Collaboration)
 } 
              

\affil{$^{1}$Code 667, NASA Goddard Space Flight Center, Greenbelt, MD 20771, USA;    \\ Email: {\tt david.bennett@nasa.gov}}
\affil{$^{2}$Department of Astronomy, University of Maryland, College Park, MD 20742, USA}
\affil{$^{3}$Warsaw University Observatory, Al.~Ujazdowskie~4, 00-478~Warszawa,Poland}
\affil{$^{4}$Department of Physics, Chungbuk National University, Cheongju 361-763, Republic of Korea}
\affil{$^{5}$Institute of Natural and Mathematical Sciences, Massey University, Auckland 0745, New Zealand}
\affil{$^{6}$Institut d'Astrophysique de Paris, 98 bis bd Arago, 75014 Paris, France}
\affil{$^{7}$Dept.\ of Astronomy, Ohio State University, 140 West 18th Avenue, Columbus, OH 43210, USA}
\affil{$^{8}$Department of Earth and Space Science, Graduate School of Science, Osaka University, Toyonaka, Osaka 560-0043, Japan}
\affil{$^{9}$Institute for Space-Earth Environmental Research, Nagoya University, Nagoya 464-8601, Japan}
\affil{$^{10}$Department of Physics, University of Auckland, Private Bag 92019, Auckland, New Zealand}
\affil{$^{11}$Okayama Astrophysical Observatory, National Astronomical Observatory of Japan, 3037-5 Honjo, Kamogata, Asakuchi, Okayama 719-0232, Japan}
\affil{$^{12}$Nagano National College of Technology, Nagano 381-8550, Japan}
\affil{$^{13}$Department of Astronomy and Astrophysics, University of California, Santa Cruz, CA 95064, USA}
\affil{$^{14}$Tokyo Metropolitan College of Aeronautics, Tokyo 116-8523, Japan}
\affil{$^{15}$School of Chemical and Physical Sciences, Victoria University, Wellington, New Zealand}
\affil{$^{16}$Institute of Space and Astronautical Science, Japan Aerospace Exploration Agency, Kanagawa 252-5210, Japan}
\affil{$^{17}$University of Canterbury Mt.\ John Observatory, P.O. Box 56, Lake Tekapo 8770, New Zealand}
\affil{$^{18}$Department of Physics, Texas A\&M University, 4242 TAMU, College Station, TX 77843-4242, USA}
\affil{$^{19}$Max-Planck-Institute for Astronomy, K\"onigstuhl 17, 69117 Heidelberg, Germany}
\affil{$^{20}$Korea Astronomy and Space Science Institute, Daejon 34055, Republic of Korea}
\affil{$^{21}$Harvard-Smithsonian Center for Astrophysics, 60 Garden Street, Cambridge, MA 02138 USA}
\affil{$^{22}$University of Canterbury, Dept. of Physics and Astronomy, Private Bag 4800, 8020 Christchurch, New Zealand}
\affil{$^{23}$Las Cumbres Observatory Global Telescope Network, 6740 Cortona Drive, suite 102, Goleta, CA 93117, USA}
\affil{$^{24}$Dept. of Earth, Atmospheric and Planetary Sciences, Massachusetts Institute of Technology, 77 Massachusetts Ave., Cambridge, MA 02139, USA}
\affil{$^{25}$ESO Vitacura, Alonso de C—rdova 3107. Vitacura, Casilla 19001, Santiago 19, Chile}
\affil{$^{26}$McDonald Observatory, 82 Mt Locke Rd, McDonald Obs TX 79734 USA}
\affil{$^{27}$School of Math and Physics, University of Tasmania, Private Bag 37, GPO Hobart, 7001 Tasmania, Australia}
\affil{$^{28}$Department of Physics, University of Rijeka, Radmile Matej v{c}i\'{c} 2, 51000 Rijeka, Croatia}
\affil{$^{29}$Technical University of Vienna, Department of Computing, Wiedner Hauptstrasse 10, 1040 Wien, Austria}
\affil{$^{30}$CFHT Corporation, 65-1238 Mamalahoa Hwy, Kamuela, Hawaii 96743, USA}
\affil{$^{31}$IRAP, CNRS - Universit\'e de Toulouse, 14 av. E. Belin, F-31400 Toulouse, France}
\affil{$^{32}$SUPA, School of Physics \& Astronomy, University of St Andrews, North Haugh, St Andrews KY16 9SS, UK}
\affil{$^{33}$Niels Bohr Institutet, K{\o}benhavns Universitet, Juliane Maries Vej 30, 2100 K{\o}benhavn {\O}, Denmark}
\affil{$^{34}$Space Telescope Science Institute, 3700 San Martin Drive, Baltimore, MD 21218, USA}
\affil{$^{35}$Department of Earth Sciences, University of California, Riverside, CA 92521,, USA}
\affil{$^{36}$South African Astronomical Observatory, PO Box 9, Observatory 7935, South Africa}
\affil{$^{37}$Astronomisches Rechen-Institut, Zentrum f{\"u}r Astronomie der Universit{\"a}t Heidelberg (ZAH), M\"onchhofstra{\ss}e 12-14, 69120 Heidelberg, Germany}
\affil{$^{38}$Perth Observatory, Walnut Road, Bickley, Perth 6076, Australia}
\affil{$^{39}$International Centre for Radio Astronomy Research, Curtin University, Bentley, WA 6102, Australia}
\affil{$^{M}$MOA Collaboration}
\affil{$^{P}$PLANET Collaboration}
\affil{$^{O}$OGLE Collaboration}
\affil{$^{\mu}\mu$FUN Collaboration}
\affil{$^{R}$Robonet Collaboration}



\begin{abstract}
We present the analysis of microlensing event MOA-2010-BLG-117, and show
that the light curve can only be explained by the gravitational lensing of
a binary source star system by a star with a Jupiter mass ratio planet.  It was necessary
to modify standard microlensing modeling methods
to find the correct light curve solution for this binary-source,
binary-lens event. We are able to measure a strong microlensing parallax signal,
which yields the masses of the host star, $M_* = 0.58\pm 0.11 \msun$, and 
planet $m_p = 0.54\pm 0.10 M_{\rm Jup}$ at a projected star-planet separation of 
$a_\perp = 2.42\pm 0.26\,$AU, corresponding to a semi-major axis of 
$a = 2.9{+1.6\atop -0.6}\,$AU. Thus, the system resembles a half-scale
model of the Sun-Jupiter system with a half-Jupiter mass planet orbiting a half-solar
mass star at very roughly half of Jupiter's orbital distance from the Sun. 
The source stars are slightly
evolved, and by requiring them to lie on the same isochrone, we can constrain
the source to lie in the near side of the bulge
at a distance of $D_S = 6.9 \pm 0.7\,$kpc, which implies a distance to the
planetary lens system of $D_L = 3.5\pm 0.4\,$kpc. The ability to
model unusual planetary microlensing events, like this one, will be necessary to 
extract precise statistical information from the planned large exoplanet microlensing
surveys, such as the WFIRST microlensing survey.
\end{abstract}


\section{Introduction}
\label{sec-intro}
Gravitational microlensing has a unique niche among planet discovery methods
\citep{bennett_rev,gaudi_araa} because of its sensitivity to planets with masses
extending to below an Earth-mass \citep{bennett96} orbiting beyond the snow
line \citep{mao91,gouldloeb92}, where planet formation is thought to be the most efficient,
according to the leading core accretion theory of planet formation \citep{lissauer_araa,pollack96}.
While radial velocity and planetary transit surveys 
\citep{wright_gaudi_book2013,kepler_q17,ida05,lecar_snowline,kennedy-searth,kennedy_snowline,thommes08} 
have found hundreds and thousands of planets, respectively,
these methods have much higher sensitivity to planets that orbit very close to their
host stars. Their sensitivity to planets like those in our own Solar System is quite
limited. Our knowledge of these wide orbit planets extending down to low masses depends on
the results of microlensing surveys \citep{suzuki16,cassan12,gould10}
This is the main reason for the selection of the space-based exoplanet microlensing survey
\citep{bennett02} as a part of the WFIRST mission \citep{WFIRST_AFTA}, which was the
top-rated large space mission in the 2010 New Worlds, New Horizons decadal survey.

Like the Kepler transit survey \citep{borucki11}, the WFIRST exoplanet microlensing survey
will primarily be a statistical survey with thousands of expected exoplanet discoveries.
However, a large number of planet discoveries does not automatically translate into good
statistics if a large fraction of the planet candidates don't allow precise interpretations
\citep{burke15,mullaly16}. Fortunately, the microlensing method predicts a relatively
small number of low signal-to-noise planet candidates \citep{gould04} compared to the
transit method. Nevertheless, microlensing does have the potential problem of microlensing
events that defy interpretation, and these could also add to the statistical uncertainty
in the properties of the exoplanet population that can be studied by microlensing. 

In the past two years, the analysis of several complicated microlensing events potentially
involving planets have been completed. The lens system for OGLE-2007-BLG-349
was revealed to be a circumbinary planet, rather than a 2-planet system with a single
host star \citep{bennett16}. This removed a significant uncertainty from the \citet{gould10},
\citet{cassan12} and \citet{suzuki16} statistical analyses, which included this event.
(If the 2-planet model for OGLE-2007-BLG-349 would have been correct, the 2nd planet
would have been the lowest mass ratio planet discovered by microlensing.)
Another complicated event was OGLE-2013-BLG-0723, which was originally claimed
to be a planet in a binary star system that was unusually close to the Sun
for a microlensing event \citep{udalski_ob130723}. This small distance to the lens system
was due to a large microlensing parallax signal. However, a more 
careful analysis of the data \citep{han_ob130723} indicated that the light curve was
better explained by a binary star system without a planet and a much smaller 
microlensing parallax signal. Most recently, \citet{han_ob160613} have analyzed a planet
in a binary star system, and found a somewhat ambiguous result with solutions consisting
of a planet and stellar (or brown dwarf) hosts with mass ratios ranging from 0.95 to
0.03.

In this paper, we present the analysis of microlensing event MOA-2010-BLG-117, an
event that eluded precise interpretation for several years after it was observed and
identified as a planetary microlensing event. It has a strong planetary signal, so it must be
included in the statistical analysis of MOA data \citep{suzuki16}. In fact, the basic
character of the light curve was obvious by inspection to many of the authors of this paper.
There was a clear planetary signal due to the crossing of two minor image
caustics, but detailed models did not provide a good fit. The region between these
two minor image caustics is an area of strong demagnification because the minor image
is largely destroyed in this region, but the magnification between the MOA-2010-BLG-117
was simply too large. It could only be fit with the addition of a fourth body to increase
the magnification between the minor image caustics. This fourth body could be a second 
source star that would not pass between the minor image caustics and would therefore
not suffer the demagnification experienced by the first source. Or the fourth body could
be a third lens that could provide additional magnification between the 
minor image caustics. We found that the only viable triple lens systems were ones
with two stars orbited by one planet, and that two planet models could not match 
the observed light curve.
The early modeling could not decide between the binary source and circumbinary planet
possibilities.

This paper is organized as follows. In Section~\ref{sec-lc_data} we describe the
light curve data, photometry and real time modeling that influenced some of the
data collection strategy. In Section~\ref{sec-lc}, we describe the systematic
light curve modeling of the final data set, which shows that the binary source model
must be correct. We also show that we can constrain the distance to the source
by requiring that the two source stars have magnitudes and colors that lie on the
same isochrone. We describe the photometric calibration and the determination
of the primary source star radius in Section~\ref{sec-radius}, and then we 
derive the lens system properties in Section~\ref{sec-lens_prop}. In
Section~\ref{sec-keck} we consider high angular resolution adaptive optics
observations of the MOA-2010-BLG-117 target, and we present a proper motion
measurement of the MOA-2010-BLG-117 target that indicates that the source star
system lies in the Galactic bulge. Out conclusions are presented in
Section~\ref{sec-conclude}.

\section{Light Curve Data, Photometry and Real Time Modeling}
\label{sec-lc_data}

Microlensing event MOA-2010-BLG-117, at ${\rm RA} =18$:07:49.67, 
${\rm  DEC} = -25$:20:40.7, and Galactic coordinates $(l, b) = (5.5875, -2.4680)$, was 
identified and announced as a microlensing candidate by the Microlensing Observations
in Astrophysics (MOA) Collaboration Alert system
\citep{bond01} on 7 April 2010. The MOA team subsequently identified the light curve
as anomalous at UT 10:19am, 2 August 2010, and this announcement triggered follow-up
observations by the Probing Lensing Anomalies NETwork (PLANET) and the MICROlensing
Follow-up Network ($\mu$FUN). The PLANET group observed this event using the 1.0m
telescope at the South African Astronomical Observatory (SAAO), and the $\mu$FUN
group used the 1.3 SMARTS telescope at the Cerro Tololo Interamerican Observatory
(CTIO).
The Optical Gravitational Lensing Experiment (OGLE) Collaboration
had just updated to their wide field-of-view OGLE-4 system \citep{ogle4}, and their 
Early Warning System (EWS) was not yet in operation with the new camera 
\citep{ogle-ews}. So, the OGLE photometry was not produced automatically by the EWS
system, but once it became clear that this event had a likely planetary signal,
OGLE began to reduce and their circulate their data.

After some systematic trends with airmass were removed from the MOA data and the
OGLE data was released, it became clear by inspection that the light curve of 
this event resembled the case of a source that crossed the region of the
triangular minor image caustics, hitting both caustics. This configuration is
somewhat similar to those of OGLE-2007-BLG-368 \citep{sumi10}
and MOA-2009-BLG-266 \citep{muraki11}, except the source for OGLE-2007-BLG-368
only crossed one of the minor image caustics and the source for MOA-2009-BLG-266
was almost as large as the minor image caustics. However, attempts to model this
event did not yield good fits with this geometry. 

\begin{figure}
\epsscale{0.9}
\plotone{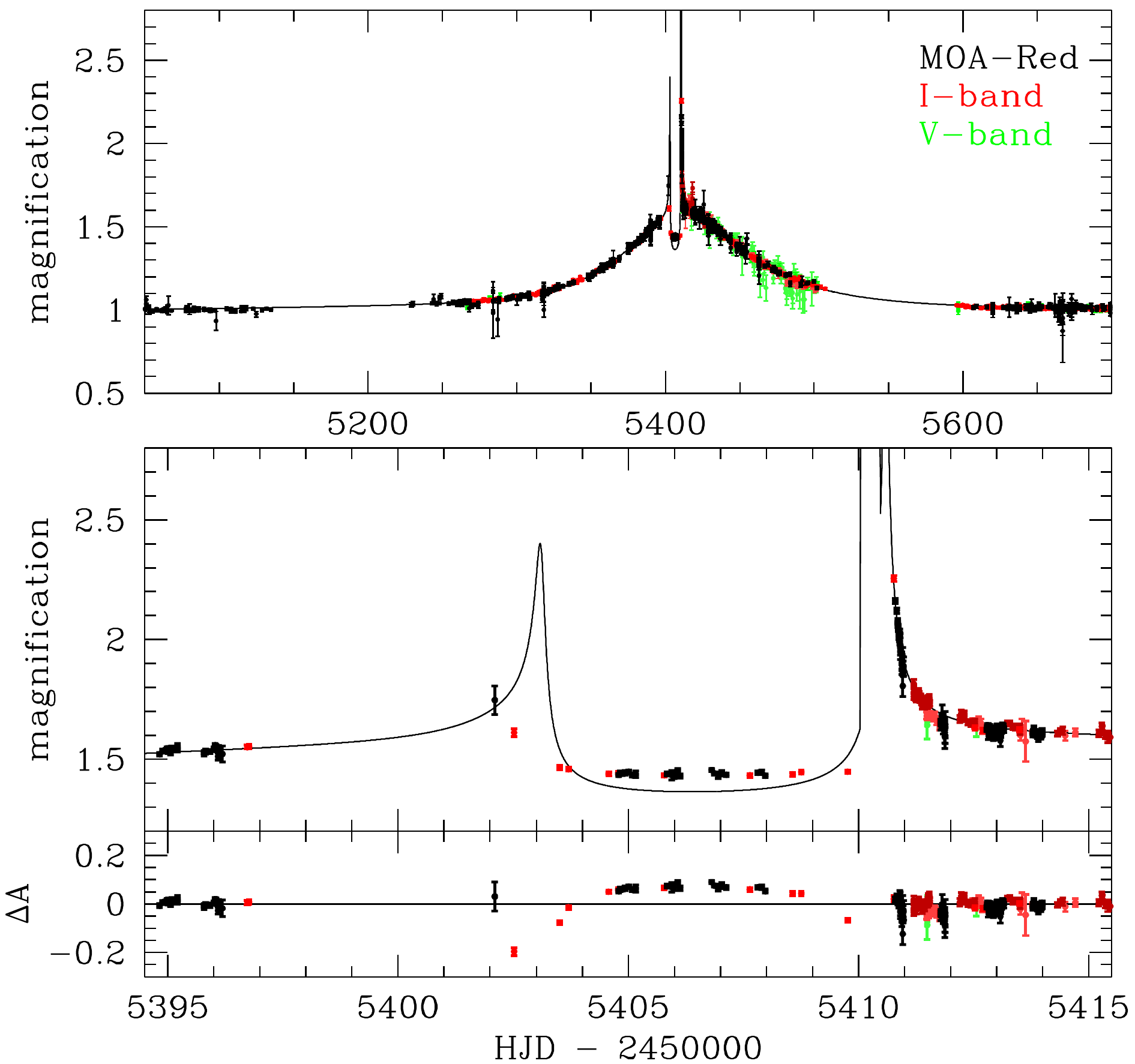}
\caption{The best binary lens model for  the MOA-2010-BLG-117 light curve.
MOA-Red band data are shown in black. $I$-band data from OGLE, CTIO, and
SAAO are shown in red, light red, and dark red, respectively, while the OGLE and
CTIO $V$-band data are shown in green and light green.
\label{fig-lcb}}
\end{figure}

The problem with this minor image caustic crossing model is that the magnification
deficit between the two caustic (or cusp) crossings at $t = 5402$ and 5411
is too small. (Note that
$t \equiv {\rm HJD} - 2450000$). This is evident in Figure~\ref{fig-lcb}, which shows
the best fit binary lens light curve for MOA-2010-BLG-2010. This light curve
has the obvious problem that the magnification between the two caustic/cusp
features is higher than the model can accommodate. In fact, the problem is
more severe than this figure indicates. In order to minimize this discrepancy
between the model and the data, the event is driven to have a very bright source,
so that the minor image will be kept at relatively low magnification, which reduces
the magnification deficit between the two caustic/cusp features. However, in this
case, the source brightness is driven to be $1.5\times$ brighter than the apparent
source star in the OGLE images. This means that negative blending is required,
since a negative "blend flux" must be added to the source flux to achieve the
relatively faint ``star" seen in the unmagnified images. Negative blending is
quite possible at low levels due to the variations in the apparent ``sky" background
due to unresolved stars, but in this case the level of negative blending is 
too large for such a physical explanation. So, it implies that this model is likely to
be incorrect.

Because of these difficulties with the minor image perturbation model
and unrelated difficulties with the real-time photometry, early
attempts at modeling this event predicted that the relatively bright, well-observed
feature at $t \approx 5411$ was the interior of a caustic entrance, where the 
caustic crossing itself was not observed. But, a subsequent caustic exit never
occurred. This made it clear that some version of a planetary minor caustic crossing 
event was correct, but that an additional lens or source was needed to explain 
the higher-than-expected brightness between the two caustic/cusp crossings.
This possibility was recognized relatively early after the discovery of the 
light curve anomaly, so we obtained more frequent CTIO $V$-band observations
than usual in the hopes that they might help reveal a color difference between
the two sources of a binary source model.

It was necessary to wait until mid-2011 before the magnification was back at baseline
because of the long duration of this microlensing event. After that,
the OGLE Collaboration provided optimal centroid photometry using the 
OGLE difference imaging pipeline\citep{ogle-pipeline}. Photometry of the 
MOA data was performed with the MOA pipeline \citep{bond01}, which
also employs the difference imaging method \citep{tom96}. The PLANET
collaboration's SAAO data was reduced with a version of the Pysis difference
imaging code \citep{albrow09}, and the CTIO data were reduced with
DoPHOT \citep{dophot}. The final data set consists of 4966 MOA observations
in the custom MOA-Red passband (roughly equivalent to the sum of
Cousins $R$+$I$), 398 and 48 OGLE observations in the $I$ and $V$ 
bands, respectively, 150 $I$-band and 88 $V$-band observations from the SMARTS telescope
in CTIO, 119 $I$-band observations from SAAO, and 10 $K$-band observations
from the VVV survey \citep{minniti-vvv} using the VISTA telescope at Paranal, which 
happened to be doing a low cadence survey of the Galactic bulge in 2010.

\section{Light Curve Models}
\label{sec-lc}

Our light curve modeling was done using the image-centered ray-shooting method
\citep{bennett96,bennett-himag}, supplemented with the hexadecapole approximation
\citep{gould_hexadec,pejcha_hexadec} that is employed when passes a test for accuracy.
For triple lens modeling, we used the code developed
for OGLE-2006-BLG-109 \citep{bennett-ogle109} and OGLE-2007-BLG-349 \citep{bennett16}.
Triple lens models have some parameters in common with single and binary lens models.
These are the Einstein radius crossing time, $t_E$, and the time, $t_0$, and distance,
$u_0$, of closest approach between the lens center-of-mass and the source star.
For a binary lens, there is also the mass ratio of the secondary to the primary lens,
$q$, the angle between the lens axis and the source trajectory, $\theta$, and the
separation between the lens masses, $s$. 

The length parameters, $u_0$ and $s$, are normalized by the Einstein radius of this total system mass, 
$R_E = \sqrt{(4GM/c^2)D_Sx(1-x)}$, where $x = D_L/D_S$ and $D_L$ and $D_S$ are
the lens and source distances, respectively. ($G$ and $c$ are the Gravitational constant
and speed of light, as usual.) For triple lens models, there are an additional separation,
mass ratio, and angle to describe the position and mass ratio of the third lens, but
we will not explore these models in detail in this paper.

For every passband, there are two parameters to describe the unlensed source
brightness and the combined brightness of any unlensed ``blend" stars that are
superimposed on the source. Such ``'blend" stars are quite common because
microlensing is only seen if the lens-source alignment is $\simlt \theta_E \sim 1\,$mas,
while stars are unresolved in ground based images if there separation is
$\simlt 1^{\prime\prime}$. However, with ground-based seeing, the background contains 
many unresolved stars, and this makes the background uneven. 
As a result, it is possible to have realistic cases of ``negative
blending" if the ``negative" brightness of the blend is consistent with the fluctuations in the 
unresolved stellar background. Artificial negative blending can occur with difference imaging photometry
that does not attempt to identify a source star in the reference image, but this
is just an artifact of the photometry method. In any case,
these source and blend fluxes are treated differently from the other parameters because
the observed brightness has a linear dependence on them, so for each set of 
nonlinear parameters, we can find the source and blend fluxes that minimize the
$\chi^2$ exactly, using standard linear algebra methods \citep{rhie_98smc1}.

For the binary source models for MOA-2010-BLG-117, we add a second source to the binary 
lens model, allowing for a different brightness and color for the second source. 
The second source has its own $t_0$ and $u_0$ values, which we denote as
$t_{0s2}$ and $u_{0s2}$. If the two source stars have exactly, the same
velocity, then the $t_E$ and $\theta$ values for the two sources would also be 
the same, but due to orbital motion, the $t_E$ and $\theta$ values are 
slightly different. However, the orbital motion of the source stars is
much smaller than the orbital motion of the source star system in the 
Galaxy, so we use parameters to describe the difference of the $t_E$ and
$\theta$ values. The parameters we use are
$dt_{Es2} = t_{Es2} -  t_{Es1}$ and $d\theta_{s2} = \theta_{s2} -\theta_{s1}$,
where $t_E = t_{Es1}$ and $\theta = \theta_{s1}$.

\begin{figure}
\epsscale{0.9}
\plotone{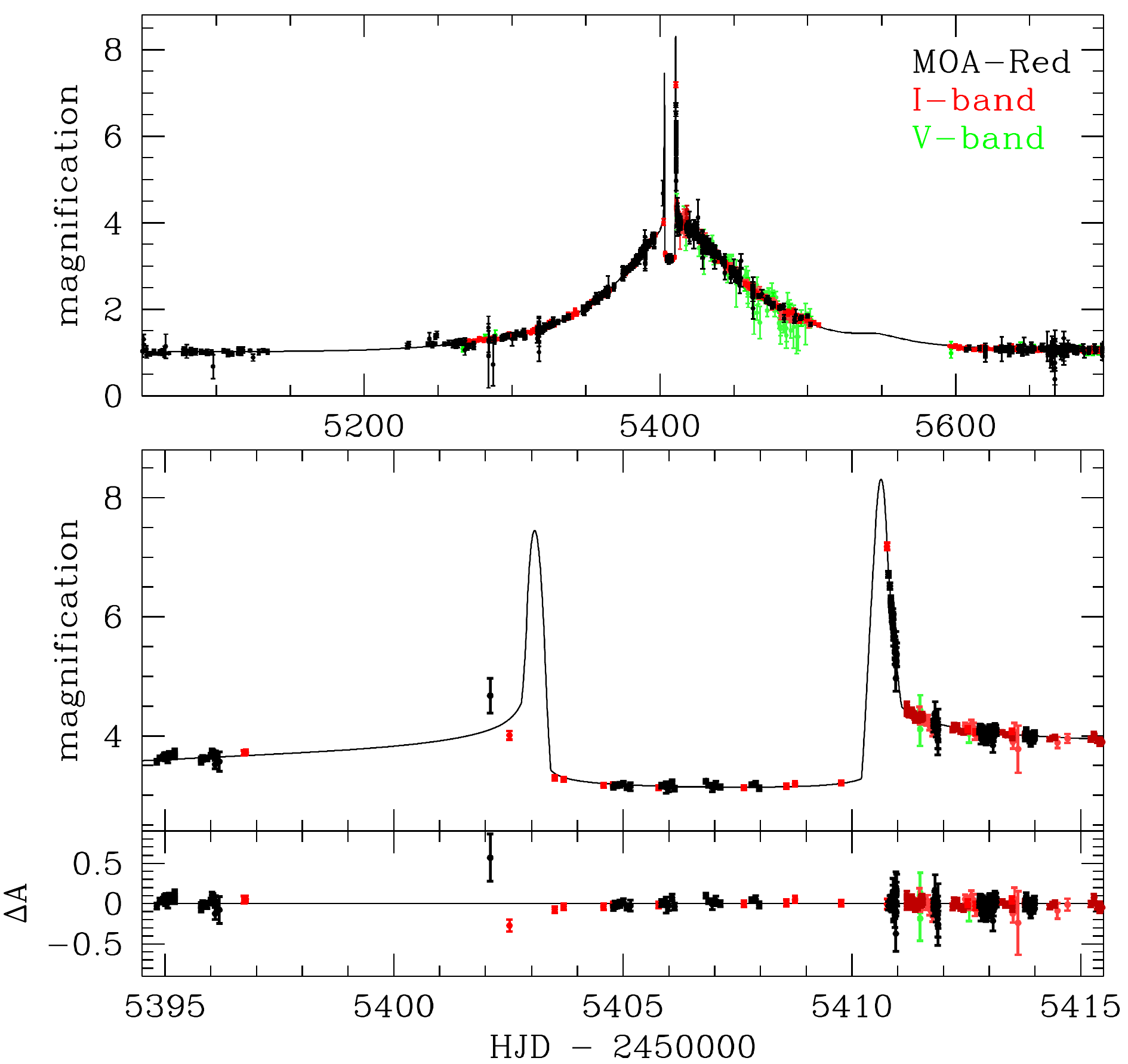}
\caption{The best circumbinary lens model for  the MOA-2010-BLG-117 light curve.
MOA-Red band data are shown in black. $I$-band data from OGLE, CTIO, and
SAAO are shown in red, light red, and dark red, respectively, while the OGLE and
CTIO $V$-band data are shown in green and light green.
\label{fig-lccb}}
\end{figure}

\begin{figure}
\epsscale{0.75}
\plotone{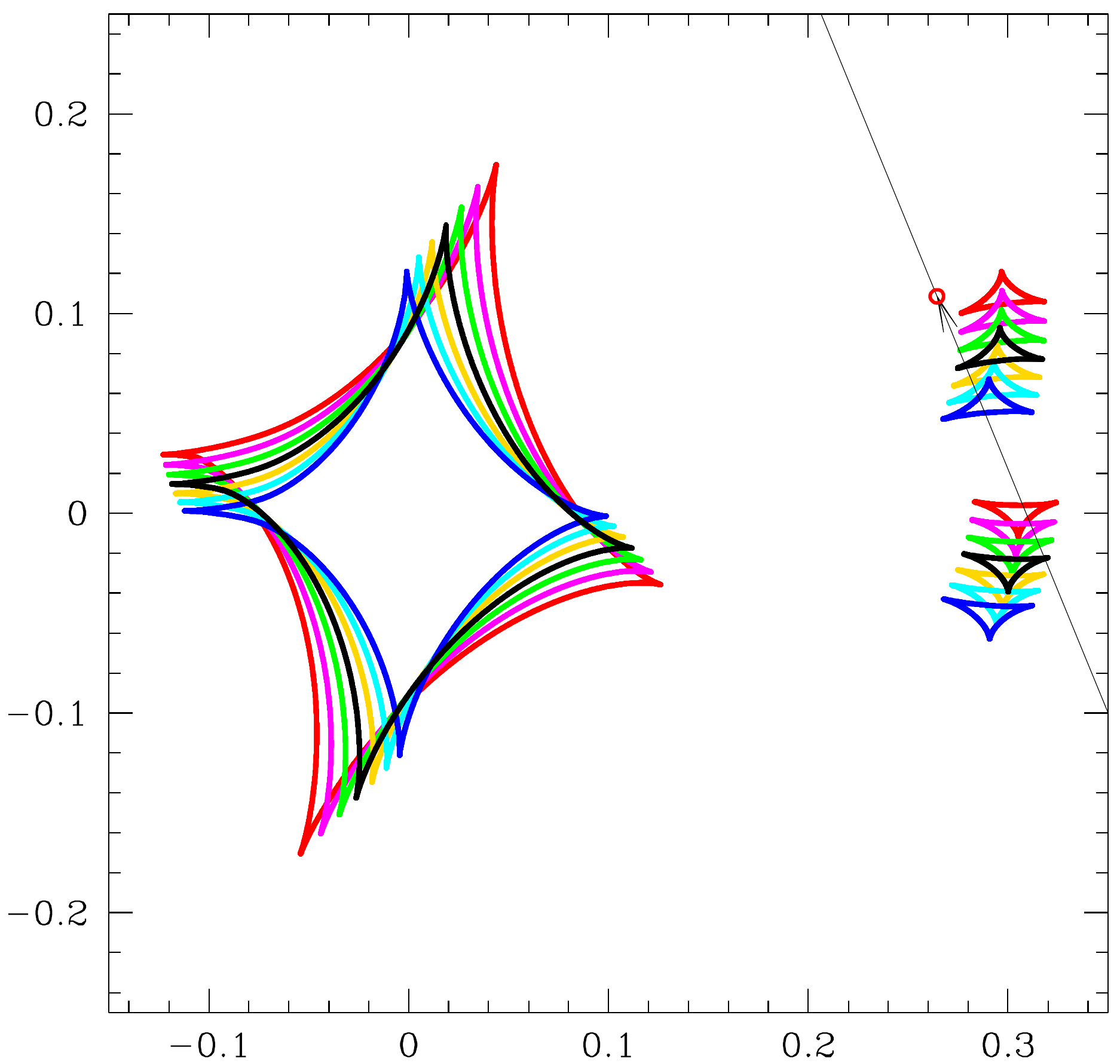}
\caption{The caustic configuration, shown at an interval of two days for the best
circumbinary lens model for  the MOA-2010-BLG-117 light curve. The caustics
are shown at $t = 5401$, 5403, 5405, 5407, 5409, 5411 and 5413 in
red, magenta, green, black, gold, cyan, and blue, respectively. The source
trajectory is given by the black line with the red circle indicating the source
size.
\label{fig-caustic_cb}}
\end{figure}

Our initial attempts to model this event favored the circumbinary models, and the
model shown in Figure~\ref{fig-lccb} was the best fit. However, there are several
problems with this model. First, although the data are sparse, the model does
not provide a good fit to the first cusp approach at $t = 5402$--5403. However,
there is a more serious problem with this model that is demonstrated by 
Figure~\ref{fig-caustic_cb}, which shows how the orbital motion of the binary
host stars affects the caustic configuration. The central caustic rotates quite
rapidly, such that the angle between the direction of the right-pointing cusp
and the source position remains nearly constant throughout the interval
between the cusp crossings. This is apparently necessary to avoid having
a local light curve peak in the middle of the long minimum at $5403.5 < t < 5410$
at a location where the cusp would be pointing directly at the source.
With the rapid orbital motion implied by this model, the source can remain at the same angle
with respect to the cusp direction throughout the passage this light
curve minimum.

The rapid orbital motion presents a problem, however. The probability of
lensing by two stars that aren't bound to each other is quite small ($\sim 10^{-12}$),
so we can assume that the two lens stars are bound. If so, then their relative
velocity can't be above the escape velocity of the system. As a result, the high 
relative velocity implies that the lens must be close to
the either the lens or the observer, because both of these possibilities allow
higher lens orbital velocities when measured in units of Einstein radii
per unit time. With the angular source radius, $\theta_*$,
derived below in Section~\ref{sec-radius}, we can derive the angular Einstein
radius, $\theta_E = \theta_* t_E/t_*$, and this yields the 
following relation \citep{bennett_rev,gaudi_araa}
\begin{equation}
M_L = {c^2\over 4G} \theta_E^2 {D_S D_L\over D_S - D_L} 
       = 0.9823\,\msun \left({\theta_E\over 1\,{\rm mas}}\right)^2\left({x\over 1-x}\right)
       \left({D_S\over 8\,{\rm kpc}}\right) \ ,
\label{eq-m_thetaE}
\end{equation}
where $x = D_L/D_S$, and $\theta_E \sim 0.8\,$mas for this event. 
This allows us to determine the lens system mass and
convert the measured transverse separation and velocity to physical units 
at every possible distance for the lens. This exercise tells us that the two stars
would be unbound for $ 0.93\,{\rm kpc} <  D_L < 7.5\,$kpc and 
$0.05\msun < M_L < 26\msun$. However, the microlensing parallax
parameters for this model imply a lens system mass of $M_L = 0.218\msun$.
We can conclude that the lens orbital velocity parameters are too large
for a physically reasonable model, so the binary source model is favored.

\begin{figure}
\epsscale{0.9}
\plotone{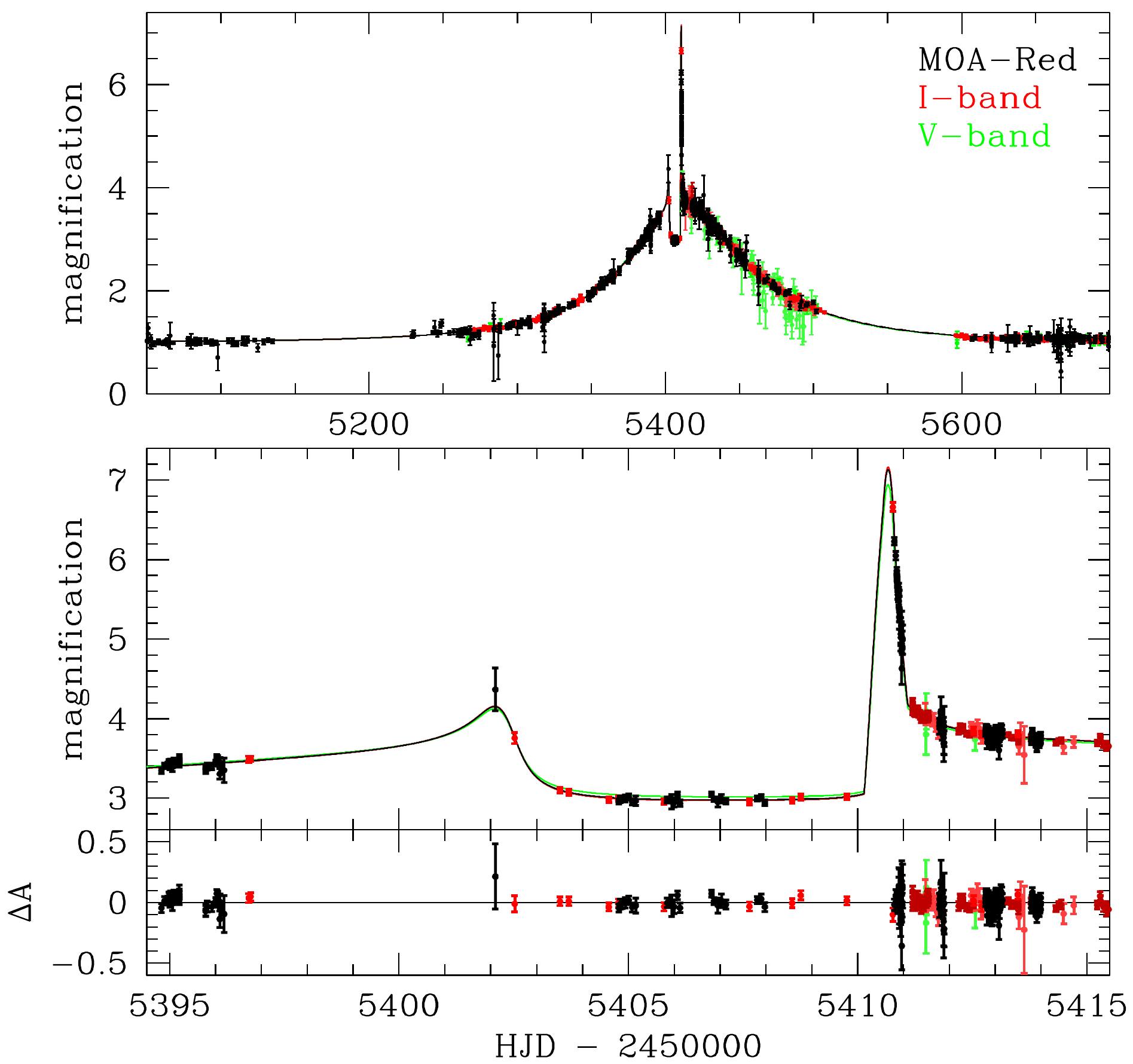}
\caption{The best binary source model for  the MOA-2010-BLG-117 light curve. The
$I$-band light curves are plotted in different shades of red, with SAAO as dark
red, OGLE as red, and CTIO as light red. The OGLE and CTIO $V$-band light curves 
are plotted in green and light green, respectively. The MOA-red band light
curve is plotted in black. The model curves for MOA-red, $I$-band, 
and $V$-band are plotted in black, red and green, respectively.
\label{fig-lc}}
\end{figure}

While the best circumbinary model implied unphysical parameters, in our initial
modeling, the best circumbinary model had a better $\chi^2$ than the best
binary source models that we found, by $\Delta\chi^2 > 130$. However, the best
binary source models from our first round of fitting had an unphysical feature,
as well. As with the models with single source, we had been considering the source
brightnesses in each passband as independent parameters. But, this allowed the
models to move into unphysical regions of parameter space, in which the flux
ratio between the two sources was very different for passbands that were nearly
identical, like the OGLE, CTIO and SAAO $I$-bands. 
In order to avoid these unphysical models, we have modified our modeling code
to fix the source flux ratio
to be the same for each of the $I$-band data sets and each of the $V$-band data sets.
The flux ratio of source-2 to source-1 is given by the parameters $f_{s2V}$ and $f_{s2I}$ in the $V$ 
and $I$-bands, respectively.
Source-1 is defined to be source that crosses the planetary caustics. For the 
MOA-red band, we do not use a independent flux ratio parameter. Instead, we derive
the MOA-red band flux ratio parameter from the $I$ and $V$-band parameters,
$f_{s2Rm} = f_{s2I}^{0.837}f_{s2V}^{0.163}$. This follows from the 
color transformation that we have derived from the bright stars in this field 
\citep{gould_col,bennett12},
\begin{equation}
R_{\rm moa} - I_{\rm O4} = 0.1630(V_{\rm O4}-I_{\rm O4}) + {\rm const} \ ,
\label{eq-Rmoa}
\end{equation}
where $V_{\rm O4}$ and $I_{\rm O4}$ refer to the OGLE-IV $V$-band and 
$I$-band magnitudes that have been used for the OGLE light curve data. Note that
these restrictions are more restrictive than those used for some previous
non-planetary binary source events that only constrained that data sets using the
same passband have the same flux ratio \citep{hwang13,jung17}.

With these limitations on the source brightness ratios, we found that the binary
source models quickly converged to a solution that was better than the previous
best binary source model by $\Delta\chi^2 \sim 200$. It was also better than
the best circumbinary model by $\Delta\chi^2 = 68.9$, even though we allowed
some of the parameters of the best circumbinary model to take unphysical values.

\begin{figure}
\epsscale{1.0}
\plottwo{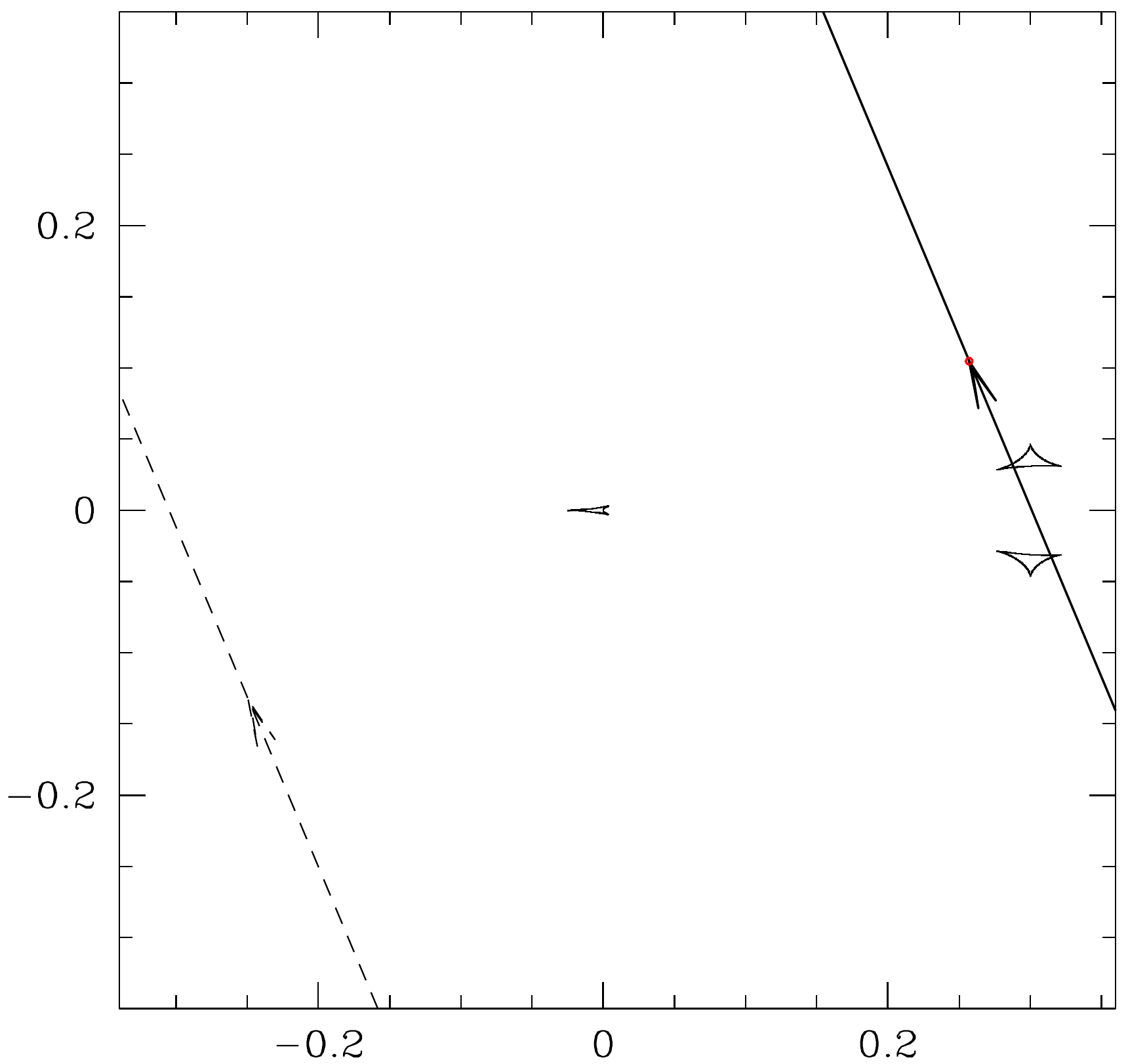}{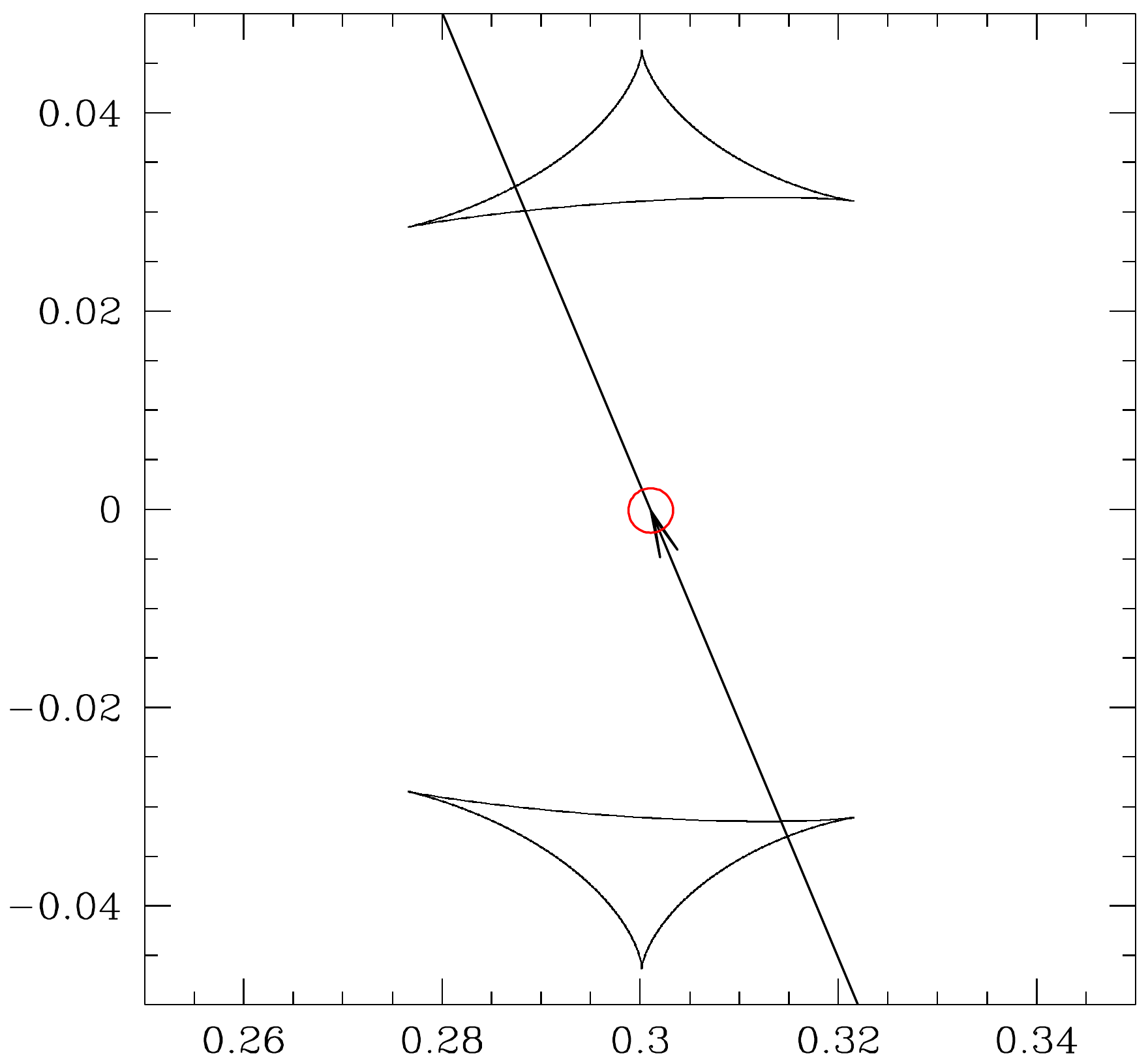}
\caption{The MOA-2010-BLG-117 caustic configuration with the source trajectories
shown as the solid and dashed curves for sources 1 and 2, respectively. The arrows
give the direction of motion for the sources with respect to the lens system,
and the red circle indicates the source of source star 1.
\label{fig-caustic}}
\end{figure}

The best fit light curve model is shown in Figure~\ref{fig-lc}, with the
parameters listed in the third column of Table~\ref{tab-mparams}. (The
best fit solution with $u_0 < 0$ is listed in the fourth column.)
Because the sources
have different colors, the light curves in the different passbands are different.
The green, red, and black curves represent the model light curves in the $V$, $I$, and
$R_{\rm moa}$ passbands, respectively. The data are plotted with a similar
color scheme. We use green and light-green for the OGLE and CTIO $V$-band data,
black for the $R_{\rm moa}$ data, and dark red, red, and light red for 
the SAAO, OGLE, and $\mu$FUN $I$-band data, respectively. The caustic configuration
for the best fit model is shown in Figure~\ref{fig-caustic}.
We define the source that crosses the 
planetary caustic to be source number 1 and the other source to be source 2.
Although both sources have similar $|u_0|\sim 0.3$ and $|u_{0s2}| \sim 0.3$ values, 
we know that only
one source comes close to the planetary caustics since we see no evidence of
a second encounter of the planetary caustics. This implies that the two sources 
must pass on different sides of the planetary host star so that the signs of
$u_0$ and $u_{0s2}$ must be different. 

\begin{deluxetable}{cccc}
\tablecaption{Best Fit Model Parameters
                         \label{tab-mparams} }
\tablewidth{0pt}
\tablehead{
\colhead{parameter}  & \colhead{units} &
\colhead{$u_0>0$} & \colhead{$u_0<0$} 
}  

\startdata
$t_E$ & days &                                124.57 & 116.64  \\
$t_0$ & ${\rm HJD}-2455400$ &    19.6850 & 19.8235  \\
$u_0$ & &                                       0.26539 & -0.29109   \\
$s$ & &                                           0.86614 & 0.85531  \\
$\theta$ & radians &                       1.95765 & -1.96029   \\
$q$ & $10^{-3}$ &                           0.8100 & 0.9451   \\
$t_\ast$ & days &                            0.3184  & 0.3511  \\
$\pi_{E,N}$ & &                        -0.1759 & 0.1916  \\
$\pi_{E,E}$ & &                         -0.0196 & -0.0394  \\
$t_{0s2}$ & ${\rm HJD}-2455400$ & 0.0189 & 0.0228  \\
$u_{0s2}$ & &                                 -0.27603 & 0.31192  \\
$f_{s2I}$ & &                                     0.7620 & 0.7631  \\
$f_{s2V}$ & &                                    0.8583 & 0.8364  \\
$dt_{Es2}$ & days &                         -9.16 & -11.80  \\
$d\theta_{s2}$ & radians &              0.32205 & -0.23631  \\
$\omega$ & $10^{-3}$ days$^{-1}$ &   -0.401   &  -1.401    \\
$\dot{s}$ & $10^{-3}$ days$^{-1}$ & -1.862 & -1.607   \\
$1/T_{S\rm orb}$ & $10^{-3}$ days$^{-1}$ & 0.5126 & 0.2016  \\
$\theta_E$ & mas &                         0.885 & 0.781  \\
fit $\chi^2$ &  &                                  5744.35 & 5748.56   \\
\enddata
\end{deluxetable}

\begin{deluxetable}{cccc}
\tablecaption{MCMC Parameter Distributions
                         \label{tab-mcmcparams} }
\tablewidth{0pt}
\tablehead{
\colhead{parameter}  & \colhead{units} & \colhead{$u_0>0$} & \colhead{$u_0<0$} 
}  

\startdata
$t_E$ & days &                                120.6(5.2) & 116.4(4.3)  \\
$t_0$ & ${\rm HJD}-2455400$ &    19.80(30) & 19.77(28)  \\
$u_0$ & &                                       0.279(15) & -0.287(13)   \\
$s$ & &                                           0.8601(60) & 0.8566(51)  \\
$\theta$ & radians &                       1.963(11) & -1.964(10)   \\
$q$ & $10^{-3}$ &                           0.950(33) & 0.952(33)   \\
$t_\ast$ & days &                            0.361(30) & 0.372(29)   \\
$\pi_{E,N}$ & &                        -0.171(20) & 0.188(22)  \\
$\pi_{E,E}$ & &                         -0.022(8) & -0.040(10)  \\
$\pi_{E}$ & &                               0.172(21) & 0.192(23) \\
$t_{0s2}$ & ${\rm HJD}-2455400$ & -0.08(60) & 0.22(9)  \\
$u_{0s2}$ & &                                 -0.267(19) & 0.301(19)  \\
$f_{s2I}$ & &                                     0.663(96) & 0.731(92)  \\
$f_{s2V}$ & &                                    0.801(117) & 0.877(113)  \\
$dt_{Es2}$ & days &                         -1.2(11.6) & -8.2(8.5)  \\
$d\theta_{s2}$ & radians &              0.305(35) & -0.300(39)  \\
$\omega$ & $10^{-3}$ days$^{-1}$ &   0.79(1.21) &  -1.17(1.17)   \\
$\dot{s}$ & $10^{-3}$ days$^{-1}$ & -1.76(30) & -1.76(30)  \\
$1/T_{S\rm orb}$ & $10^{-3}$ days$^{-1}$ & 0.50(13) & 0.49(12)  \\
$\theta_E$ & mas &                         0.805(100) & 0.777(95)  \\
$\mu_{\rm rel,G}$ & mas/yr &           2.46(31) & 2.44(31)  \\
$V_{s1}$ & &                                          20.43(7) & 20.38(6)   \\
$I_{s1}$ & &                                           17.95(7) & 17.90(6)   \\
$K_{s1}$ & &                                           14.90(8) & 14.85(7)  \\
$V_{s2}$ & &                                          20.69(11) & 20.53(10)   \\
$I_{s2}$ & &                                           18.41(11) & 18.25(10)   \\
$K_{s2}$ & &                                           15.58(14) & 15.42(14)   \\
$K_{s12}$ & &                                           14.43(4) & 14.34(4)   \\

\enddata
\end{deluxetable}

The model parameters for the best fit models with $u_0 > 0$ and $u_0 < 0$
are given in Table~\ref{tab-mparams}. 
Table~\ref{tab-mcmcparams} gives the Markov Chain
Monte Carlo (MCMC) averages for the models parameters. This table also
includes some derived parameters of physical interest: the angular Einstein
radius, $\theta_E$, the microlensing parallax amplitude, $\pi_E = \sqrt{\pi_{E,E}^2+\pi_{E,N}^2}$,
and the lens-source relative proper motion, $\mu_{\rm rel,G}$, in a inertial geocentric frame
that moves with the Earth at time $t_{\rm fix} = 5411\,$days.
The source-lens relative velocities for the two sources should be approximately
equal because orbital velocity of two stars separated by approximately an
Einstein radius in the Galactic bulge is typically about an order of magnitude
smaller than the orbital velocity of stars in the inner Galaxy. So, we expect the
lens-source relative velocity vectors for the two sources to differ by no more than
$\sim 10$\%. However, a $\sim 10$\%
difference between the $t_E$ and $\theta$ values for the two sources will have a 
significant effect on the light curve shape, so we must include parameters to 
describe $t_E$ and $\theta$ for the second source. We chose the parameters
$dt_{Es2} \equiv t_{Es2} - t_{Es1}$, where $t_{Es1} \equiv t_E$ and $t_{Es2}$
are the $t_E$ values for the two sources. The different source trajectory angle
is described by $d\theta_{s2} \equiv \theta_{s2} - \theta_{s1}$, where $\theta_{s1} \equiv \theta$
and $\theta_{s2}$ are the angles between the source trajectories and the lens
axis. We also allow for orbital acceleration of the two source stars. We assume a
circular orbit for these stars with an orbital period of $T_{S\rm orb}$ and projected
velocities at time $t_{\rm fix} = 5411\,$days implied by the $dt_{Es2}$ and $d\theta_{s2}$
values. These are circular orbits in three dimensions following the parameterization
of \citet{bennett-ogle109}.

The orbital velocities in the lens system are also important, but since the planetary
features in the light curve are detectable for only $\sim 10\,$days, we do not need to
include the orbital acceleration of the source. We describe the lens orbital velocities
with a rotation of the lens system with angular frequency $\omega$ and a velocity
of $\dot{s}$ in the separation direction.

This event has a significant
orbital microlensing parallax signal \citep{gould-par1,macho-par1}, with a $\chi^2$ improvement
of $\Delta\chi^2 = 43.93$ with nearly equal contributions from the MOA and OGLE
data sets. The microlensing parallax is defined by a two dimensional vector, $\piEbold$ with
North and East components of $\pi_{E,N}$ and $\pi_{E,E}$ in a geocentric
coordinate system moving with the velocity of the Earth measured at time $t_{\rm fix} = 5411\,$days.
The parameter $t_{\rm fix}$ is also the reference time for the source and lens positions.

We should note that there are a upper limits on the relative velocities between the 
two sources and between the lens star and planet since they must (almost certainly)
be gravitationally bound systems.
We assume that the source stars each have a solar mass
and compare the 2-dimensional kinetic energy to the maximum binding energy of
the source stars (using their separation on the plane of the sky). Then, following
\citet{muraki11}, we apply a constraint on the $dt_{Es2}$ and $d\theta_{s2}$ 
values. For the lens system, we know the lens mass from the microlensing parallax
parameters and the angular Einstein radius, $\theta_E$ \citep{gould-par1,bennett_rev,gaudi_araa},
and we use this to apply the same constraint. In both cases, the orbital semi-major axis
is proportional to $\theta_E$.

\begin{figure}
\epsscale{0.9}
\plotone{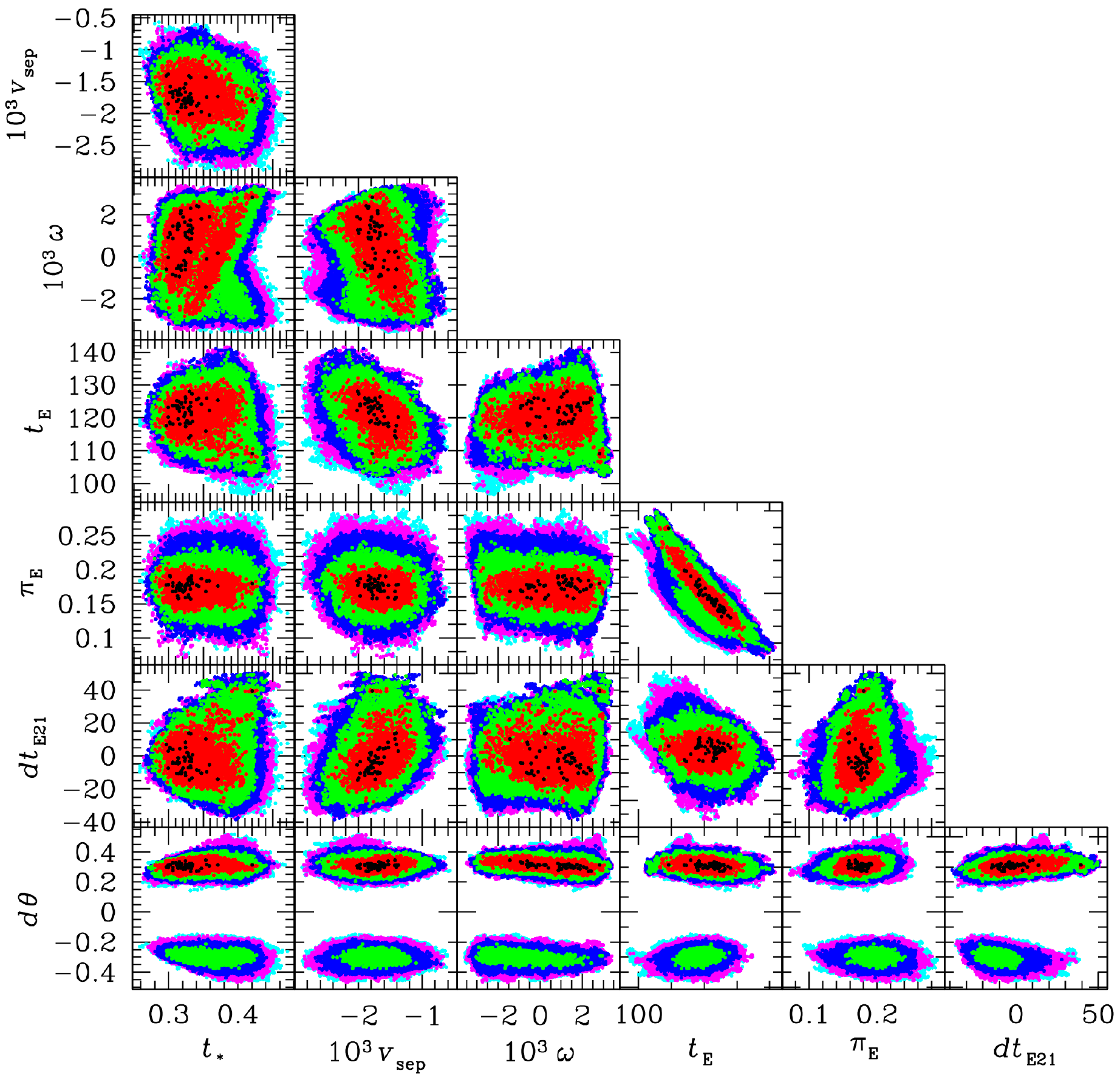}
\caption{Correlations from our MCMC runs between lens velocity parameters ($v_{\rm sep}$
and $\omega$) and the parameters that affect the inferred host star mass:
$d\theta$, $t_*$, $t_E$, $\pi_E$, and $dt_{E21}$, from our MCMC runs. 
Smaller $t_*$ values imply larger $\theta_E$ values,
which implies tighter constraints on the parameters that describe the source velocities,
$d\theta$, and $dt_{E21}$. Black, red, green,
blue,magenta, and cyan indicate models that have $\chi^2$ values larger than the
best fit model by $\Delta\chi^2 < 1$, 
$1 < \Delta\chi^2 < 4$, $4 < \Delta\chi^2 < 9$, $9 < \Delta\chi^2 < 16$, 
$16 < \Delta\chi^2$, and $\Delta\chi^2 > 16$ respectively. 
\label{fig-param_cor}}
\end{figure}

These lens and source orbital motion constraints are sensitive to the source radius crossing
time through $\theta_E = t_E \theta_*/t_*$, but the light curve constraint on $t_*$ is relatively
weak because the caustic crossings are only partially covered.
The initial fits to this event with no microlensing parallax, no lens orbital motion,
and $dt_{Es2} \equiv 0$ and $d\theta_{s2} \equiv 0$ had a large
variation in $t_*$ values ranging from $0.24\,$days to $0.40\,$days. 
When we allowed
the $dt_{Es2}$ and $d\theta_{s2}$ values to vary, subject only to the constraint on the 
maximum orbital motion of the source stars, we found that large values of these parameters
were preferred. However, the semi-major axis of the orbit
of the source stars is proportional to $\theta_E = t_E \theta_*/t_*$. Thus, a larger
$t_*$ implies a smaller $\theta_E$ and therefore a smaller semi-major axis. The
smaller semi-major axis implies a higher gravitational binding energy, which allows larger lens
star velocities implying larger values for $dt_{Es2}$ and $d\theta_{s2}$.
Since the data apparently prefer larger values for $dt_{Es2}$ and $d\theta_{s2}$, the
constraint on $t_*$ becomes tighter when we include non-zero values of $dt_{Es2}$ and $d\theta_{s2}$
and apply the orbital motion constraint. This can be seen from Figure~\ref{fig-param_cor}.
Values of $t_* < 0.26\,$days are excluded and the 2-$\sigma$ lower limit on $t_*$ is $t_* > 0.30\,$days.
Also, large values of $|dt_{Es2}|$ and $|d\theta_{s2}|$ are excluded for the smallest $t_*$
values. The microlensing parallax amplitude, $\pi_E$, is not strongly correlated with any
of the source or lens orbital motion parameters. It does have a strong anti-correlation with
the Einstein radius crossing time, but this is just a well known feature of the blending degeneracy
that is responsible for the uncertainty in $t_E$.

The $\chi^2$ difference between the $u_0 > 0$ and $u_0 <0$ solutions is small, as
indicated in the bottom row of Table~\ref{tab-mparams}. The 
$u_0 > 0$ solution is best, with the best $u_0 <0$ solution disfavored by 
$\Delta\chi^2 = 4.21$. This small
$\chi^2$ differences imply that all of these solutions will contribute to the physical parameter
probability distributions, but the $u_0 > 0$ solutions will dominate.

\begin{figure}
\epsscale{0.9}
\plotone{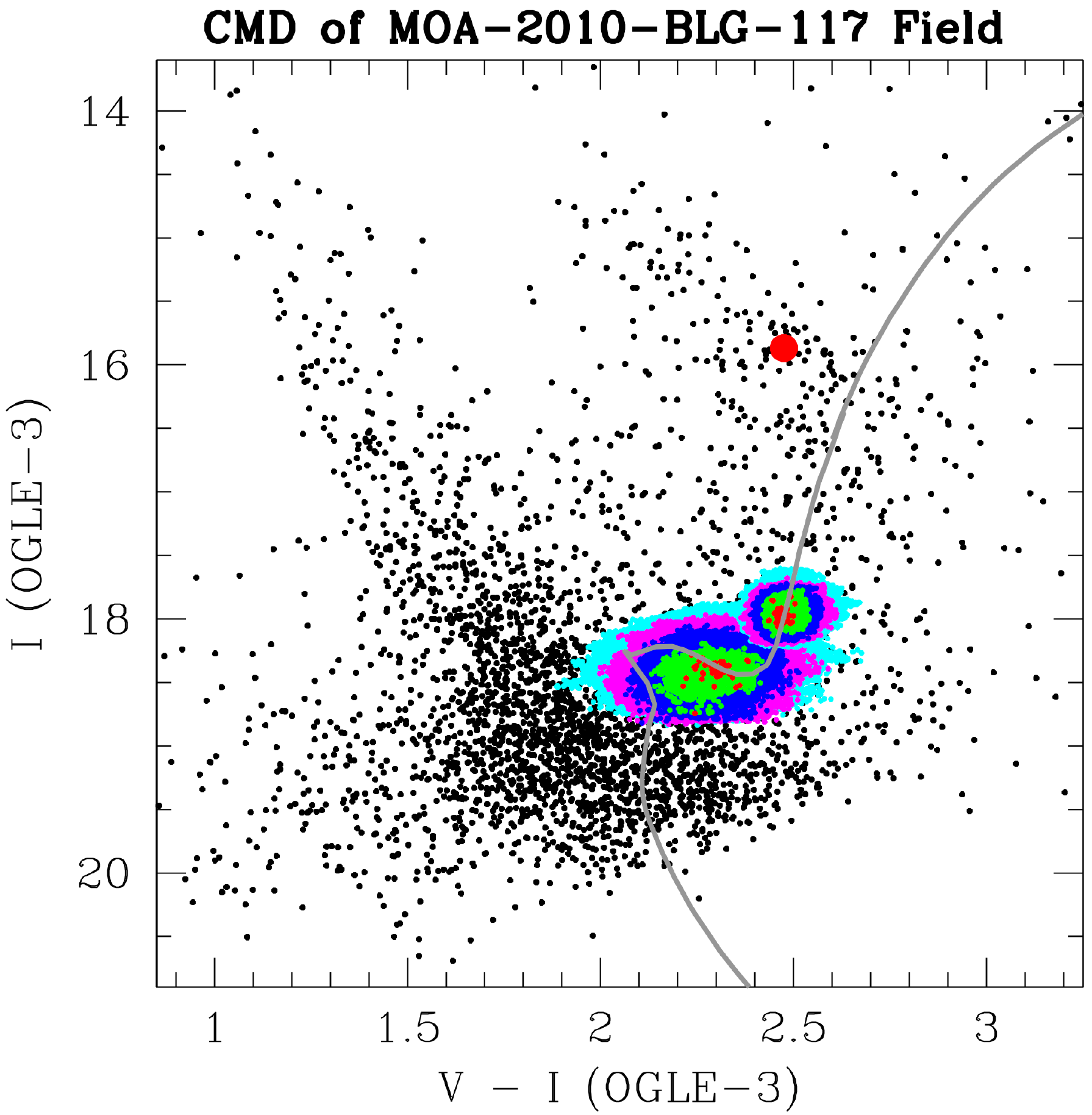}
\caption{The $(V-I,I)$ color magnitude diagram (CMD) of the stars in the OGLE-III catalog
\citep{ogle3-phot}
within $90^{\prime\prime}$ of MOA-2010-BLG-117. The red spot indicates
red clump giant centroid, and the smaller spots of different colors indicate the 
magnitude and colors of the two sources from our MCMC calculations. Red, green,
blue,magenta, and cyan indicate models that have $\chi^2$ values larger than the
best fit model by $\Delta\chi^2 < 1$, 
$1 < \Delta\chi^2 < 4$, $4 < \Delta\chi^2 < 9$, $9 < \Delta\chi^2 < 16$, and
$16 < \Delta\chi^2$, respectively. Source 1 is brighter and redder than
source 2 for most models. The grey line indicates the isochrone that best matches the
source magnitudes and colors of the best fit model. This isochrone has an age of
$4.0\,$Gyr and a metalicity of $[{\rm Fe/H}] = 0.28$.
\label{fig-cmd}}
\end{figure}

An unusual feature of this event is that the source system consists of two stars that have both
left the main sequence. Contrary to the situation for main sequence stars, the fainter
star is bluer than the brighter star for most of the solutions that comprise our Markov chains.
This can be seen from Table~\ref{tab-mcmcparams} and even more clearly in
the color magnitude diagram shown in 
Figure~\ref{fig-cmd}. This will allow us to constrain the source distance by requiring
that the source stars lie on the same isochrone in Section~\ref{sec-lens_prop}.

\section{Photometric Calibration and Primary Source Radius}
\label{sec-radius}


In order to measure the angular Einstein radius, $\theta_E = \theta_\ast t_E/t_\ast$, 
we must determine the angular radius of the source star, $\theta_\ast$,
from the dereddened brightness and color of the 
source star \citep{kervella_dwarf,boyajian14}. We determine the calibrated source
brightness in the $V$ and $I$-bands by calibrating the OGLE-IV light curve
photometry to the OGLE-III catalog \citep{ogle3-phot}. This gives:
\begin{eqnarray}
V_{\rm O3cal} &= 0.2643 + V_{\rm O4} - 0.0855\left(V_{\rm O4}- I_{\rm O4}\right)
\label{eq-V-cal} \\
I_{\rm O3cal} &= 0.0403 + I_{\rm O4} + 0.0032\left(V_{\rm O4}- I_{\rm O4}\right)
\label{eq-I-cal} \ ,
\end{eqnarray}
where $V_{\rm O4}$ and $I_{\rm O4}$ are the OGLE-IV light curve magnitudes and
$V_{\rm O3cal} $ and $I_{\rm O3cal} $ are the calibrated OGLE-III magnitudes. 


In order to estimate the source radius, we need extinction-corrected magnitudes,
and we determine these from the magnitudes and colors of the centroid of the
red clump giant feature in the OGLE-III color magnitude diagram (CMD), as indicated
in Figure~\ref{fig-cmd}. Using the red clump centroid finding method of
\citet{bennett-ogle109}, we find the red clump centroid to be located at 
$I_{\rm O3rc} = 15.868$ and $V_{\rm O3rc}-I_{\rm O3rc} = 2.475$. We compare
this to the predicted extinction corrected red clump centroid magnitude and color
of $I_{\rm rc0} = 14.288$ and $V_{\rm rc0} - I_{\rm rc0} = 1.06$, which is
appropriate \citep{nataf13,bensby13} for the Galactic coordinates 
of this event, $(l,b) = (5.5875,-2.4680)$. This yields extinction values of 
$A_I = 1.580$ and $A_V = 2.995$, which implies an extinction corrected primary source
magnitude and color of $I_{s1,0} = 16.421$ and $V_{s1,0} - I_{s1,0} = 1.052$
for the best fit model.

These dereddened magnitudes can be used to determine the angular source radius,
$\theta_*$. We use the relation from the analysis of \citet{boyajian14}, but with a
restricted range of colors corresponding to $3900<T_{\rm eff}<7000$
(Boyajian, private communication, 2014). We use
\begin{equation}
\log_{10}\left[2\theta_*/(1 {\rm mas})\right] = 0.501414 + 0.419685\,(V-I)_{s1,0} -0.2\,I_{s1,0} \ ,
\label{eq-theta_star} 
\end{equation}
and this gives $\theta_* = 2.20\,\mu$as for the best fit model. Now, there is some indication
of differential reddening in the CMD (Figure~\ref{fig-cmd}), so this can add some uncertainty
to our determination of $\theta_*$. Fortunately the effect of this uncertainty in the extinction
tends to cancel contributions from $(V-I)_{s1,0}$ and $I_{s1,0}$ in equation~\ref{eq-theta_star}.
To account for this uncertainty, we add 13\% uncertainty to our extinction estimates, which
translates into a 9\% uncertainty in $\theta_*$, according to equation~\ref{eq-theta_star},
to be used in our MCMC calculations.
As Figure~\ref{fig-cmd} indicates, the uncertainty
in the magnitude and color of source 1 is larger than the uncertainty for most events.
This is because flux can be traded between the two sources. However, this
source radius determination is correlated with the other microlens model parameters,
particularly the Einstein radius crossing time, $t_E$, which occurs in the $\theta_E = \theta_* t_E/t_*$
formula. Therefore, we determine $\theta_E$ for each model in our MCMC, and this yields
the $\theta_E$ values listed in Table~\ref{tab-mcmcparams}: $\theta_E = 0.805 \pm 0.100\,$mas
for the $u_0 > 0$ solutions and $\theta_E = 0.777 \pm 0.095\,$mas
for the $u_0 < 0$ solutions.

\begin{deluxetable}{cccc}
\tablecaption{Physical Parameters\label{tab-pparam}}
\tablewidth{0pt}
\tablehead{
\colhead{Parameter}  & \colhead{units} & \colhead{value} & \colhead{2-$\sigma$ range} }
\startdata 
$D_S $ & kpc & $6.9\pm 0.7$ & 5.6-8.3 \\
$D_L $ & kpc & $3.5\pm 0.4$ & 2.9-4.3 \\
$M_h$ & $\msun$ & $0.58 \pm 0.11$ & 0.39-0.83 \\
$m_p$ & $M_{\rm Jup}$ & $0.54\pm 0.10$ & 0.38-0.77 \\
$a_\perp$ & AU & $2.42\pm 0.26$ & 1.93-2.97  \\
$a_{3d}$ & AU & $2.9{+1.6\atop -0.6}$ & 2.1-10.3  \\
$V_L$ & mag & $24.3^{+1.5}_{-1.7}$ & 21.1-27.2 \\
$I_L$ & mag & $21.2^{+1.0}_{-1.1}$ & 19.1-23.0 \\
$K_L$ & mag & $18.3^{+0.6}_{-0.8}$ & 16.7-19.6 \\
\enddata
\tablecomments{ Uncertainties are
1-$\sigma$ parameter ranges. }
\end{deluxetable}

\begin{figure}
\epsscale{0.7}
\plotone{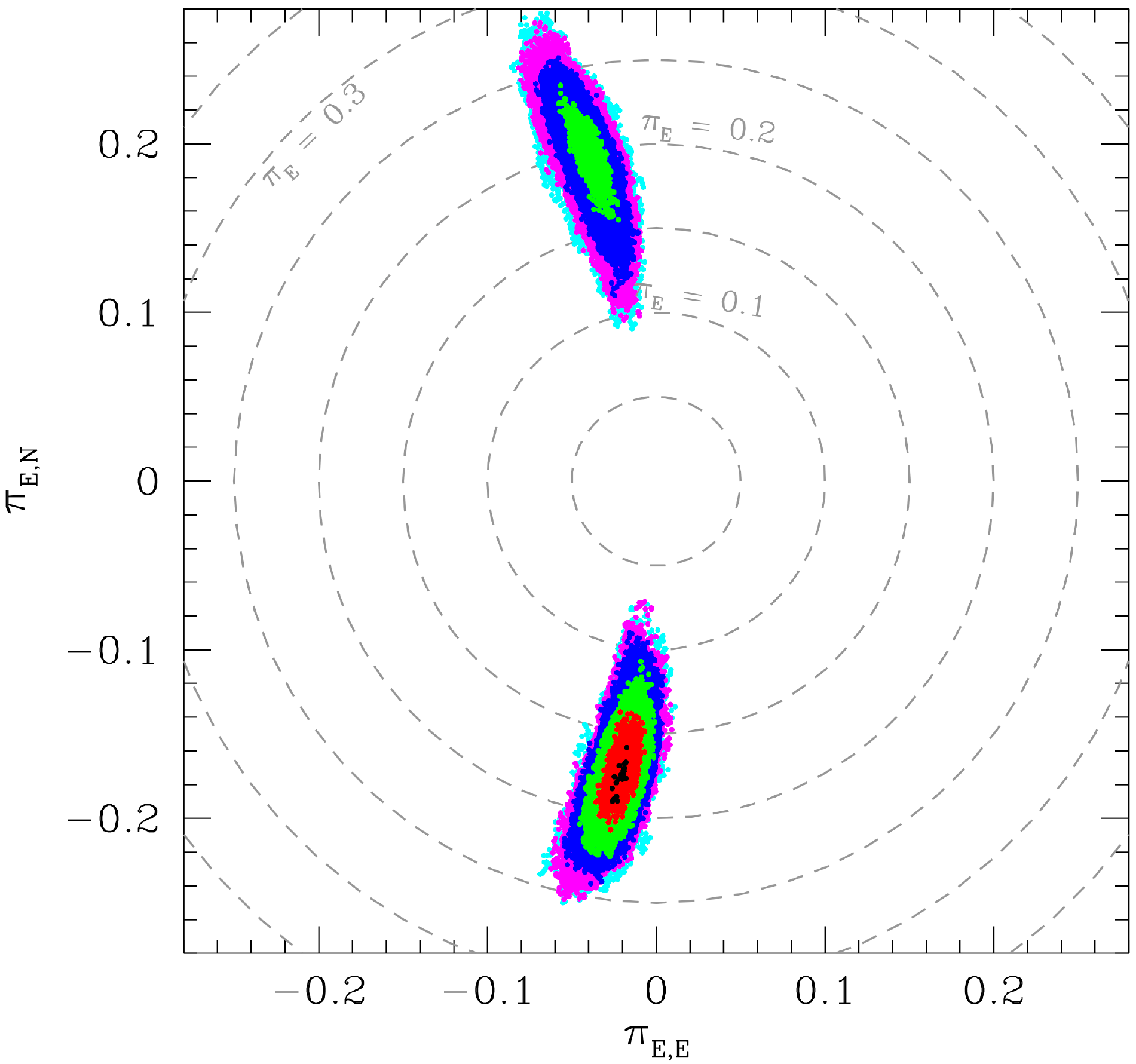}
\caption{The values of the microlensing parallax vector, $\piEbold$, from our MCMC
runs are shown. The $u_0 > 0$ solutions have $\pi_{E,N} < 0$ and are preferred
over the $u_0 < 0$ solutions (with $\pi_{E,N} > 0$) by $\Delta\chi^2 = 9.17$. The
MCMC points are color-coded. The points within $\Delta\chi^2 < 1$ are black, and
the points within $1 < \Delta\chi^2 < 4$, $4 < \Delta\chi^2 < 9$, $9 < \Delta\chi^2 < 16$,
$16 < \Delta\chi^2 < 25$, and $25 < \Delta\chi^2$ are red, green, blue, magenta,
and cyan, respectively. The dashed circles indicate curves of constant $\pi_{\rm E}$.
\label{fig-piE}}
\end{figure}

\section{Lens System Properties}
\label{sec-lens_prop}

When both the angular Einstein radius, $\theta_E$, and the microlensing parallax,
$\piEbold$, are measured, we can use
the following relation \citep{gould-par1,an-eros2000blg5,gould-lmc5},
\begin{equation}
M_L = {\theta_E c^2 {\rm AU}\over 4G \pi_E} = {\theta_E \over (8.1439\,{\rm mas})\pi_E} \msun \ ,
\label{eq-m}
\end{equation}
to determine the mass of the lens system, but in our case, we have degenerate solutions
to consider. The degeneracy allowing different $t_*$ values is probably unique to the specific
circumstances of this event. However, the degeneracy between the $u_0 > 0$ and $u_0<0$ solutions
is a very common degeneracy due to the reflection of the lens plane with respect to the
orientation of the Earth's orbit, which allows us to measure the parallax effect
with ground-based data. For high magnification
events, the lens-source system has an approximate reflection symmetry, so this
$u_0 > 0 \leftrightarrow u_0<0$ degeneracy has little effect on $\pi_E \equiv |\piEbold |$.
Because the binary source system for MOA-2010-BLG-117 has $u_0 \approx -u_{0s2}$
and source 2 is only $\sim 0.3\,$mag fainter than source 1, the lens and source 
system in this event also has an approximate symmetry (assuming that the planetary
feature has little influence on the microlensing parallax signal). This could be the
reason why the distributions of the $\piEbold$ vector, shown in Figure~\ref{fig-piE}
also show this approximate reflection symmetry.  This figure shows the distributions
from both degenerate solutions with best fit parameters listed in Table~\ref{tab-mparams}
and Markov chain distributions listed in Table~\ref{tab-mcmcparams}. The $u_0 > 0$ and 
$u_0 < 0$ solutions
are widely separated with opposite signs for the $\pi_{E,N}$ values. These opposite
signs mean that the $|\pi_{E,N}|$ values are very similar for all solutions. The $\pi_{E,E}$
values are also similar and much smaller than $|\pi_{E,N}|$, so the $\pi_E$ values
for all the degenerate solutions are similar. This means that there is overlap
in the mass distributions predicted by all four degenerate solutions.

As mentioned in Section~\ref{sec-lc}, we impose a requirement that both sources
lie on the same isochrone. This requirement is not imposed during the light curve
modeling, but it is imposed in our Bayesian analysis that uses all the models
from our Markov chains to determine the physical parameters
of the lens system. Each light curve model in our Markov chains is weighted by 
the $\chi^2$ of the best fit of the model source magnitudes and colors
to the isochrones. Thus, the location of the source magnitudes and colors
in Figure~\ref{fig-cmd} does not depend on these isochrones, but the color coding
of the source magnitudes and colors does depend on the depend on the fit to the 
isochrones. We use isochrones from the PAdova and TRieste Stellar Evolution Code 
(PARSEC) project
\citep{bressan12_PARSEC,chen14_PARSEC,chen15_PARSEC,tang14_PARSEC}.
We find that our modeling results are consistent with isochrones with ages
in the range 4--$10\,$Gyr and metalicity in the range $ -0.04 \leq [{\rm Fe/H}]\leq 0.58$.
These values are quite typical of Galactic bulge stars, as indicated by 
the microlens source stars with high resolution
spectra taken at high magnification by \citet{bensby13,bensby17}. However, it is also
possible that the source might have slightly higher extinction than the average of the
red clump stars. In that case, the range of allowed source star metalicities might extend
to sub-solar metalicities.

The main practical effect of this isochrone constraint is to force the source star system to
be located on the near side of the bulge. The isochrones prefer a source distance of
$D_S = 6.2 \pm 1.1\,$kpc, but when this priors on the source density and microlensing
probability are included, this shifts to $D_S = 6.9 \pm 0.7\,$kpc, as given in Table~\ref{tab-pparam}. 

\begin{figure}
\epsscale{0.56}
\plotone{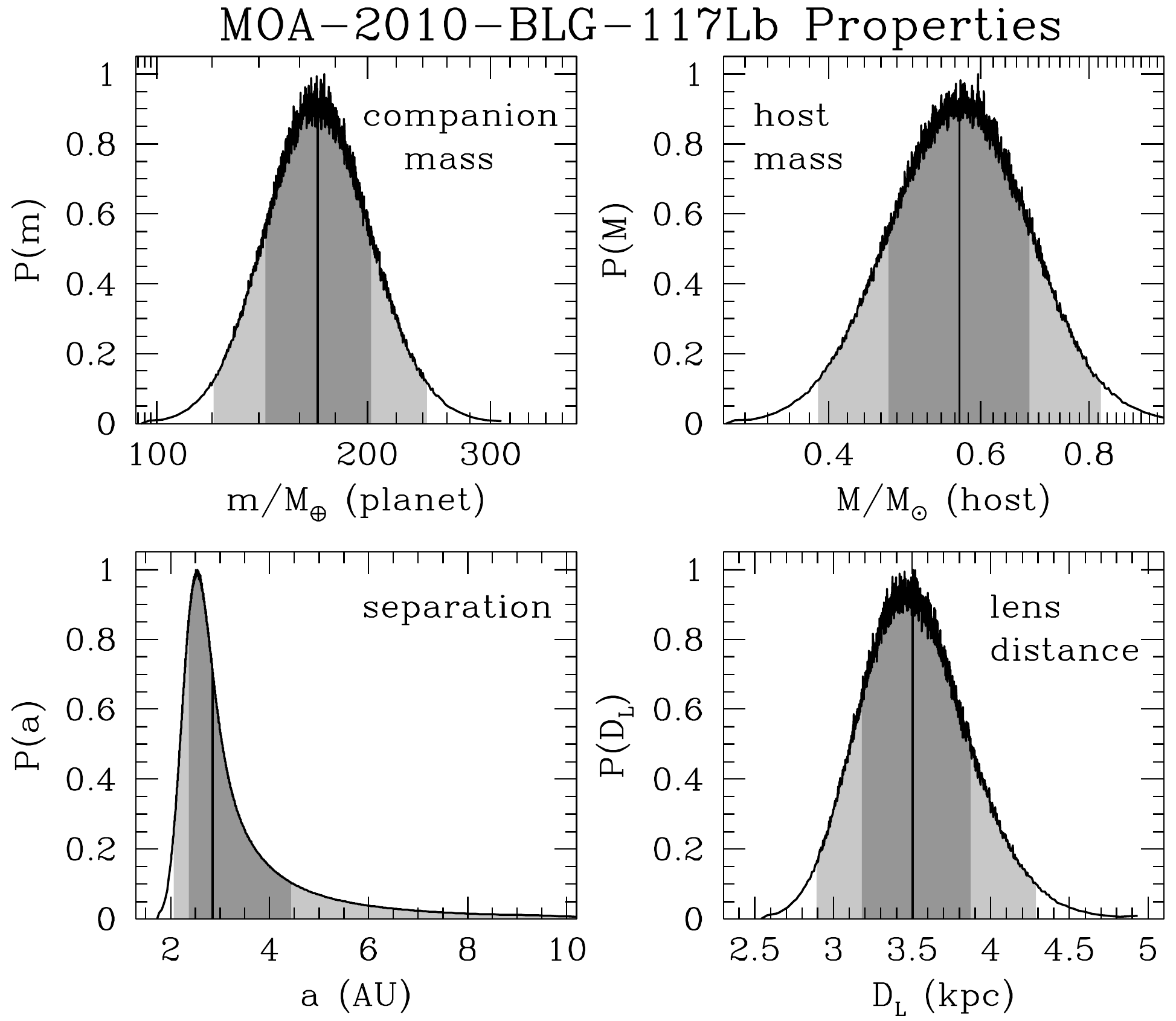}
\caption{Probability distributions of the planet and host star mass, three dimensional separation
and lens system distance based on a Bayesian analysis using mass and distance determinations
from the MCMC light curve distributions along with prior probabilities from a standard
Galactic model.
\label{fig-lens_prop}}
\end{figure}

\begin{figure}
\epsscale{0.56}
\plotone{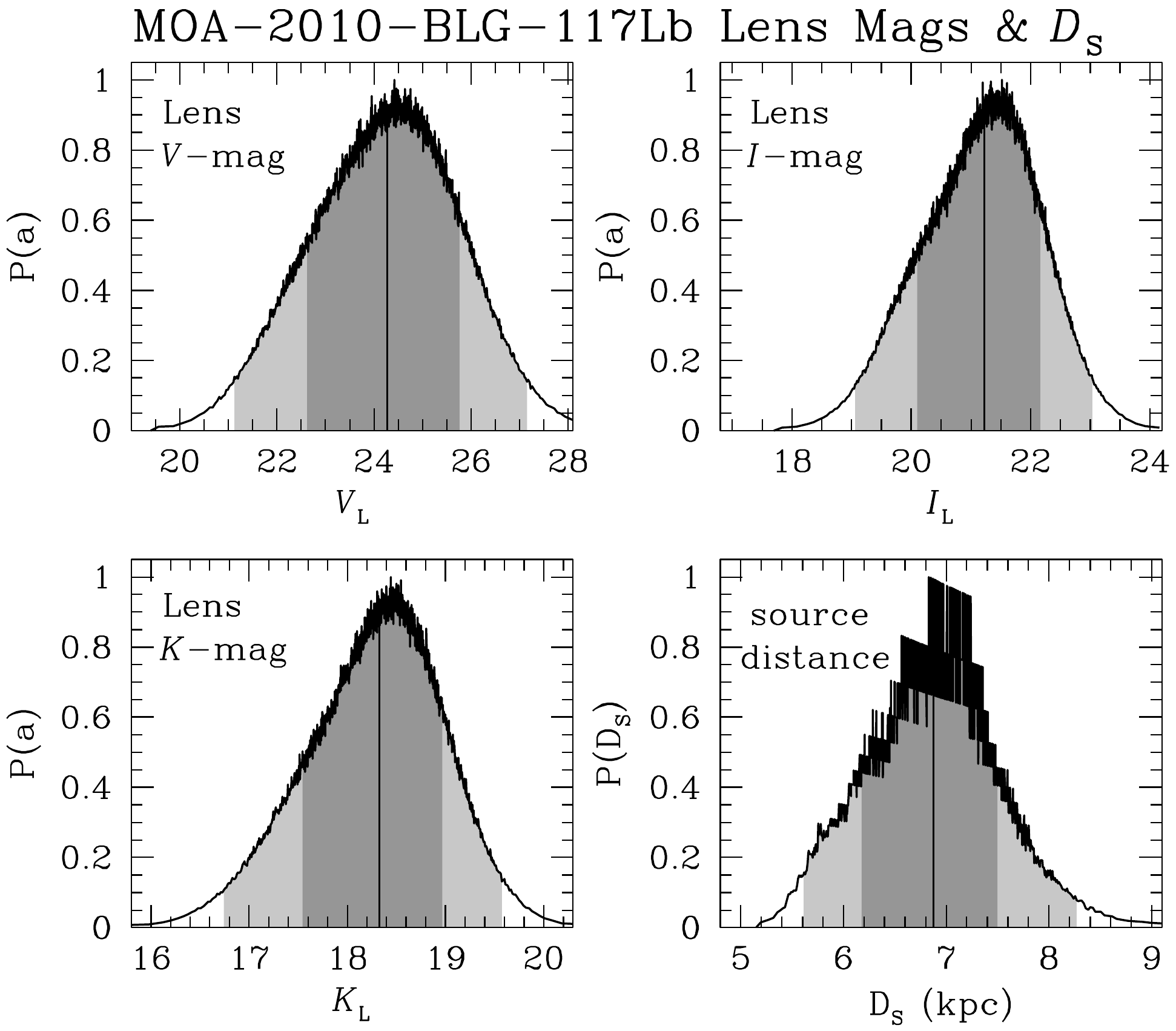}
\caption{Probability distributions of the host star $V$, $I$, and $I$ magnitudes, based on our
light curve models and Bayesian priors from a standard Galactic model. 
\label{fig-lens_prop2}}
\end{figure}

We determine the physical parameters of this lens system with a Bayesian analysis 
marginalized over the Galactic model used by \citet{bennett14}, and the results are summarized
in Figures~\ref{fig-lens_prop} and \ref{fig-lens_prop2}, as well as
Table~\ref{tab-pparam}. The host star and planet masses ($M_h$ and $m_p$) are determined
directly from equation~\ref{eq-m} with the $\pi_E$, $q$, source magnitude and color values
determined for each model in our MCMC. The $\theta_*$ and $\theta_E$ values are 
determined directly from equations~\ref{eq-theta_star}, \ref{eq-V-cal} and \ref{eq-I-cal} for
each model. The $u_0 < 0 $ solutions are weighted by $e^{-\Delta\chi^2/2} = 0.122$ with respect
to the $u_0 > 0$ solutions, where
$\Delta\chi^2 = 4.21$ is the $\chi^2$ difference between the best fit solutions with parameters
listed in Table~\ref{tab-mparams}. There is no appreciable difference in the parameter space
volume covered by the two solutions, so this approach is adequate.
The Galactic model prior has little influence on the lens mass determination
because the prior has little variation over the parameter values that are consistent with
the MCMC light curve models. The Galactic model has a larger influence on the distance
to the lens, because the stellar density has a strong dependence on 
the distance to the source star, $D_S$. The relation between the distances to the lens and source
stars is given by
\begin{equation}
D_L={{\rm AU}\over \pi_{\rm E}\theta_{\rm E}+\pi_S} \ ,
\label{eq-Dl}
\end{equation}
where $\pi_S$ is the parallax of the source star, $\pi_S = {\rm AU}/D_S$. As Table~\ref{tab-pparam}
indicates, these calculations indicate that the host star has a mass of $M_h = 0.58 \pm 0.10\msun$
and the planet has a mass of $M_p = 0.51 \pm 0.07M_{\rm Jup}$, where $M_{\rm Jup}$
is the mass of Jupiter. Assuming a random orientation, their 3-dimensional separation is
$a_{3d} = 2.9{+1.6\atop -0.5}\,$AU. The planet mass uncertainty is smaller than the
host mass uncertainty because the high host star mass ($t_* < 0.34\,$days), $u_0 > 0$ solutions have
a lower mass ratio than the other solutions, as indicated in Tables~\ref{tab-mparams} 
and \ref{tab-mcmcparams}.

The predicted host (and lens) star $V$, $I$, and $K$ magnitudes are shown in Figure~\ref{fig-lens_prop2},
along with the source distance, $D_S$. The distribution of $D_S$ favor a large number of discrete
values. This is due to our requirement that the two source stars lie on the same isochrone and
the discrete values of the metalicity, $[{\rm Fe/H}]$, 
and $\log({\rm Age})$ at intervals of 0.04 and 0.05, respectively. 

\begin{figure}
\epsscale{0.9}
\plottwo{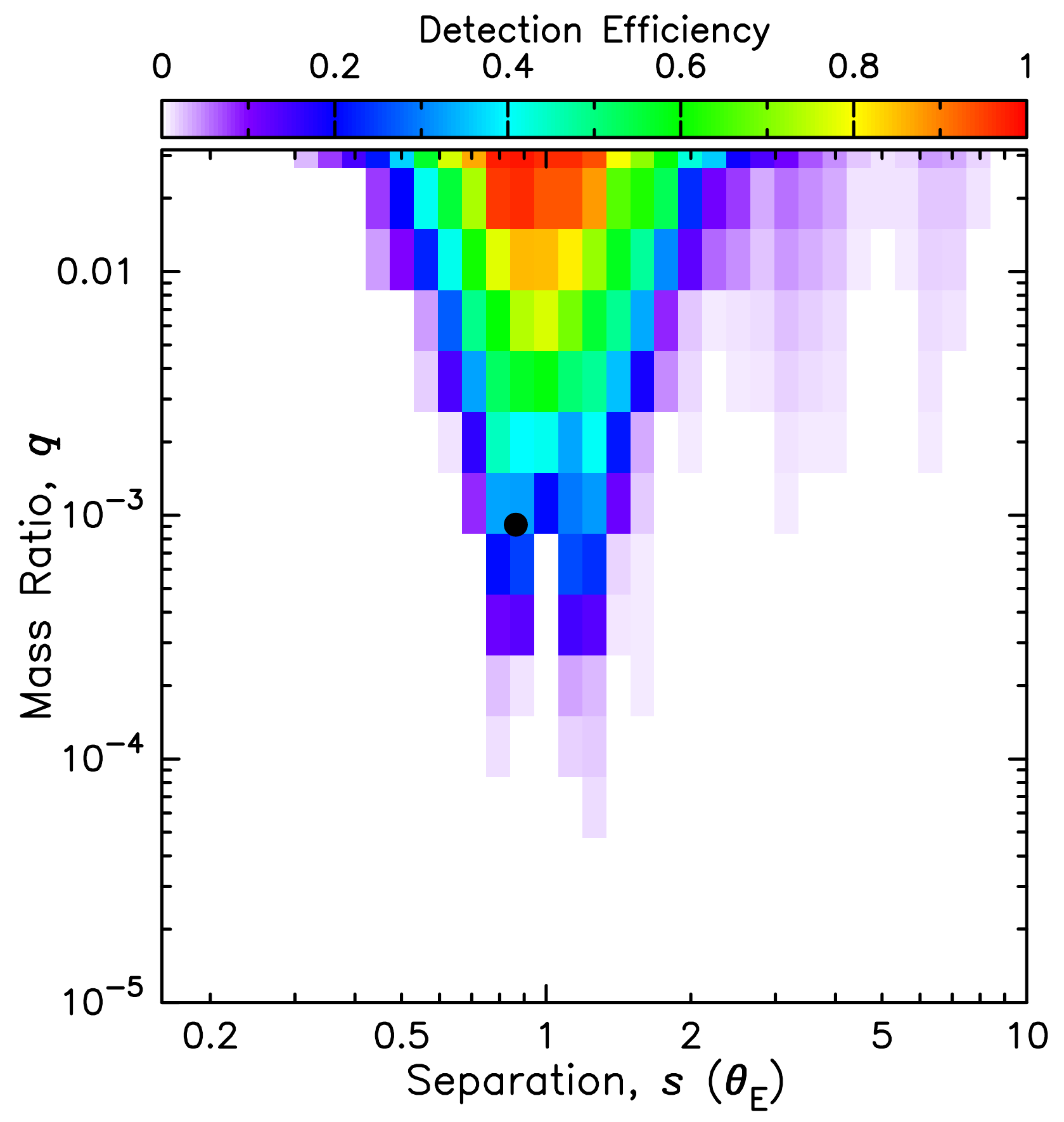}{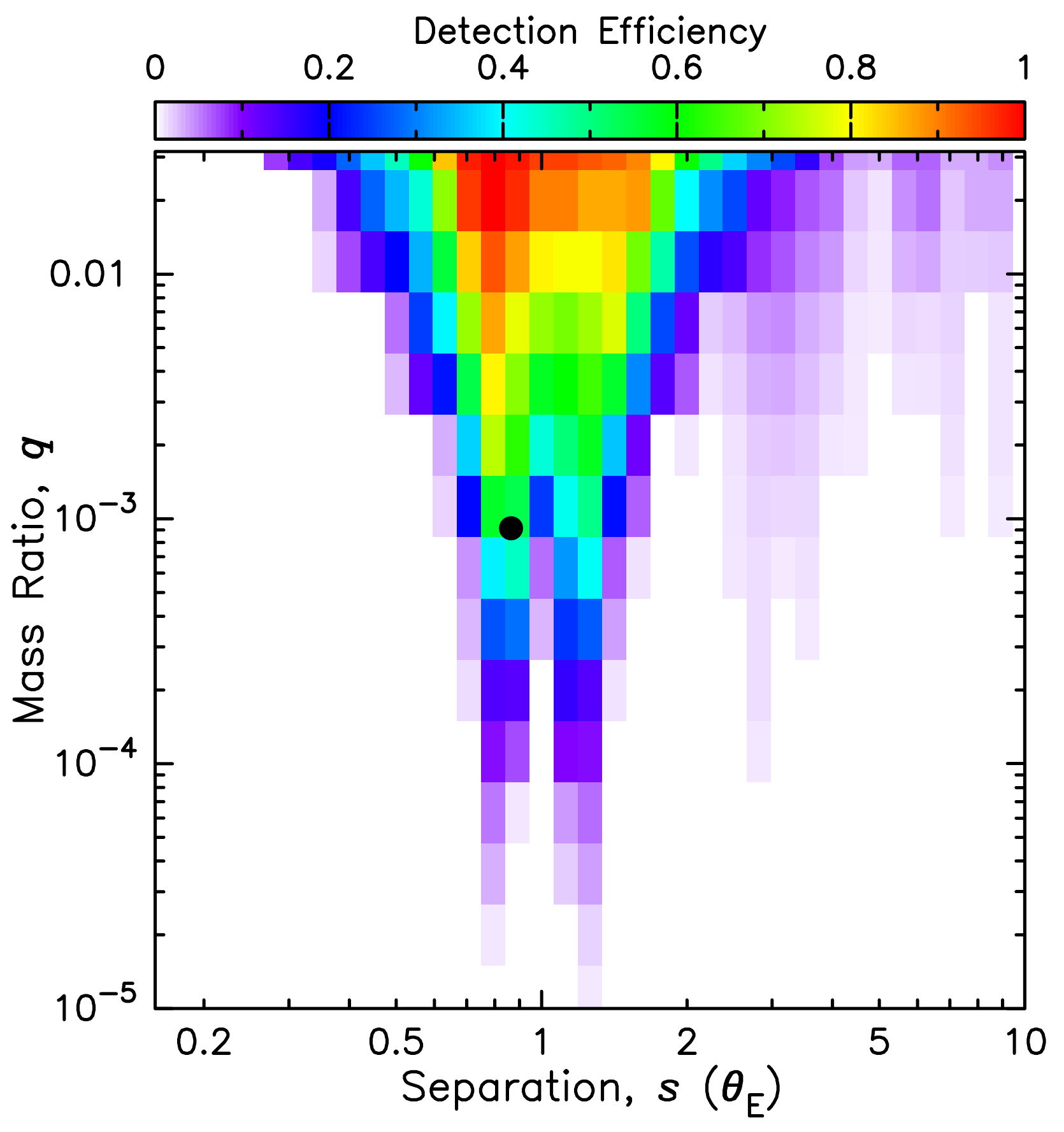}
\caption{Planetary detection efficiency for MOA-2010-BLG-117. The left panel shows the
detection efficiency due to the source star that led to the real planet detection, and
the right panel shows the planet detection efficiency for the actual event with both
source stars. In both cases, the black spots indicate the position of the planet.
\label{fig-eff}}
\end{figure}

The additional source star also increases our odds of detecting planets orbiting the
lens star because the second source provides a second probe of the lens plane.
This can be seen in Figure~\ref{fig-eff}, which shows the two cases from our
recent exoplanet mass ratio function paper \citep{suzuki16}. Over much of
the parameter range the second source star approximately
doubles the planet detection efficiency.
However, this is a much smaller increase than is provided by high magnification events.

\section{Keck Follow-up Observations}
\label{sec-keck}

\begin{figure}
\epsscale{0.7}
\plotone{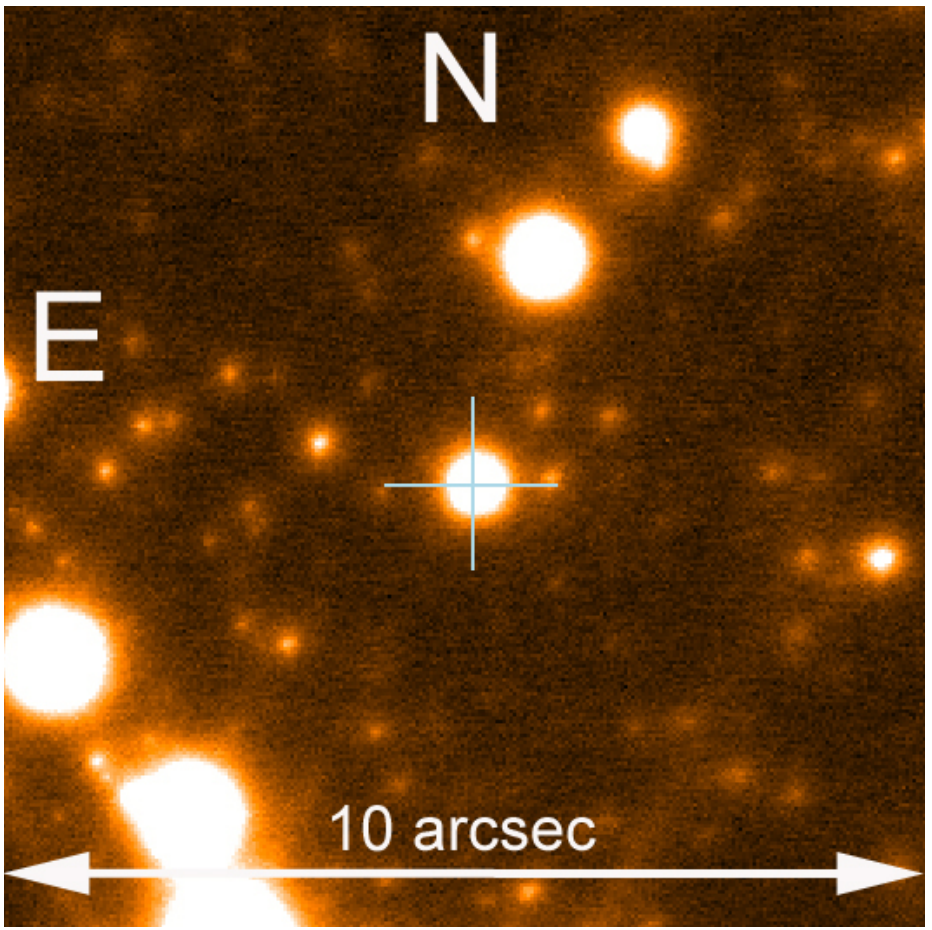}
\caption{The co-added Keck AO image of the target star is indicated by the
crosshairs.  The target consists of the 
combined flux of the source stars, the lens (and planetary host star), any
bound companions to either the source or lens system, and possibly
an unrelated star that happens to be located $\simlt 0.2^{\prime\prime}$
from the source.
\label{fig-keckim}}
\end{figure}

In an attempt to identify the lens and planetary host star, we have obtained
high angular resolution adaptive optics (AO) observations from the Keck 2 telescope.
Unfortunately, the seeing conditions were relatively poor compared to some of our
other Keck observations \citep{batista15} that achieved a point spread function
(PSF) full-width half-max (FWHM) of $60\,$mas. Our stacked $K$-band image of
the MOA-2010-BLG-117 field has a PSF FWHM of $220\,$mas, and it is shown
in Figure~\ref{fig-keckim}. The Keck images were taken in 2012, two years after
the event. With a lens-source relative proper motion of 
$\mu_{\rm rel,G} = 2.45\pm 0.31\,$mas/yr, there is no chance to detect the lens-source
separation either through image elongation \citep{bennett07,bennett15} or a color-dependent
image centroid shift \citep{bennett06}. However, there is still a chance to detect
the unresolved lens star flux on top of the flux from the source stars. In this case, the
source stars are relatively bright sub-giants, so it would be difficult to detect a host star
as faint as the star indicated by the finite source and microlensing parallax measurements,
as discussed in Section~\ref{sec-lens_prop}.

\begin{figure}
\epsscale{0.7}
\plotone{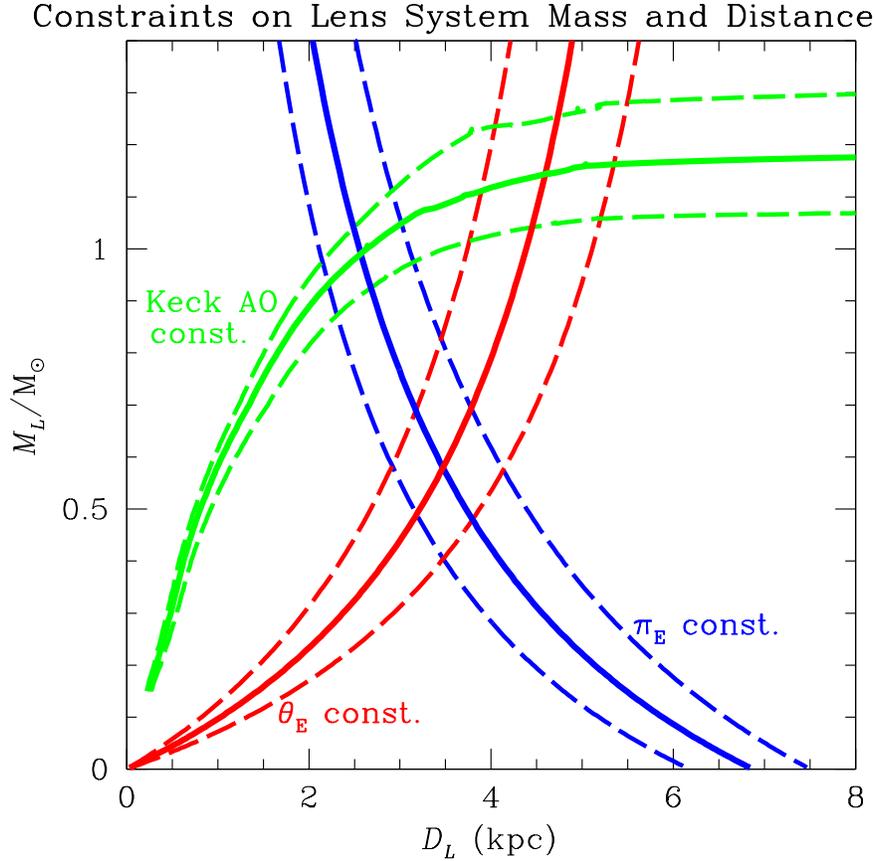}
\caption{Constraints on the host star mass and distance from the microlensing 
parallax, $\pi_E$ (in blue), the angular Einstein radius, $\theta_E$ (in red), and the host star flux
(in green), under the assumption that 
the excess flux observed in the Keck AO images is due to the host star.
The dashed line indicate approximate 1-$\sigma$ uncertainty contours (that ignore
the correlations between the parameters). For the excess $K$-band flux, the
solid green line is the from a $5\,$Gyr isochrone and the dashed green lines represent
$3\,$Gyr and $7\,$Gyr isochrones 
\citep{bressan12_PARSEC,chen14_PARSEC,chen15_PARSEC,tang14_PARSEC}.
\label{fig-2mass_const}}
\end{figure}

The ``star" detected in the Keck AO images is indeed significantly brighter,
$K_{\rm Keck} = 13.97 \pm 0.04$, than the
combined flux of the source stars, which is $K_{s12} = 14.43 \pm 0.04$.
However, this excess blend flux at $K_b = 15.12 \pm 0.15$ does not
match the lens mass and distance derived in Section~\ref{sec-lens_prop}.
The predicted host star brightness is $K_L = 18.3^{+0.6}_{-0.8}$, and as
can be seen from Figure~\ref{fig-lens_prop2}, the probability of the lens 
(and host) star being brighter than $K_L < 16$ is negligible.
Figure~\ref{fig-2mass_const} compares the constraints from the microlensing
parallax, angular Einstein radius and lens flux constraints, assuming that the
excess flux is due to the lens star. Obviously, these constraints are not
consistent with each other. The most likely solution to this inconsistency is
simply that the excess flux is not due to the lens. The other possibilities that
could explain this excess flux at the position of the source star are 
a binary companion to the lens, a tertiary companion to the source stars, or
an unrelated star. A Bayesian analysis using the measured bulge luminosity
function and measured frequencies of multiple star systems (Koshimoto \etal\
in preparation, 2017) gives similar probabilities for each of these possibilities,
with slightly larger probabilities for lens and source companions than for an
unrelated star.

While we believe that the result from the $\pi_E$ and $\theta_E$ measurements
is very likely to be the correct interpretation, we will briefly consider that one of
these measurements is wrong. From Figure~\ref{fig-2mass_const}, we see that
a host star mass of $M_h \sim 1\msun$ at a distance of $D_L \sim 2.6\,$kpc
would be favored if the blend flux is due to the lens star and the $\pi_E$ measurement
is correct. Alternatively, if the $\pi_E$ measurement was incorrect, while the $\theta_E$
measurement was correct and the blend flux is due to the lens star, then
the lens star would have to be an evolved star above a solar mass. The green
isochrone curves in Figure~\ref{fig-2mass_const} are nearly horizontal where
they cross the red $\theta_E = \,$const.\ curve. This is due to the fact that stars
evolve very quickly through these evolved phases, and this implies that this
solution is particularly unlikely.

A final possibility is that the $\pi_E$ and $\theta_E$ measurements are correct,
and the excess flux comes from the planetary host star. This would imply that the
assumption made for the red and blue $\theta_E$ and $\pi_E$ curves in 
Figure~\ref{fig-2mass_const} that the source is in the Galactic bulge 
(at $D_S = 6.8\pm 0.6\,$kpc) is not correct. From equation~\ref{eq-Dl}, we have
$ D_S = D_L/( 1 -  \pi_{\rm E}\theta_{\rm E}D_L/{\rm AU})$,  and this tells us that
if the lens system is located at $D_L \approx 0.9\,$ where the green lens flux curve crosses
the $M_h = 0.58 \pm 0.10$ value indicated by the $\theta_E$ and $\pi_E$ measurements
(according to equation~\ref{eq-m}),
then the source would be at a distance of $D_S = 1.04\,$kpc. This is highly unlikely or
at least ruled out for
two reasons. First, the rate that stars at this distance are microlensed is more than
two orders of magnitude lower than the rate that bulge stars are microlensed. 
Second, the two source stars appear to reside on the Galactic bulge sub-giant
branch of the CMD, shown in Figure~\ref{fig-cmd}. Very few foreground
disk stars to lie on this portion of the CMD, and there virtually no way to arrange
for the fainter star in a binary pair to be bluer than the brighter star.

\begin{figure}
\epsscale{0.71}
\plotone{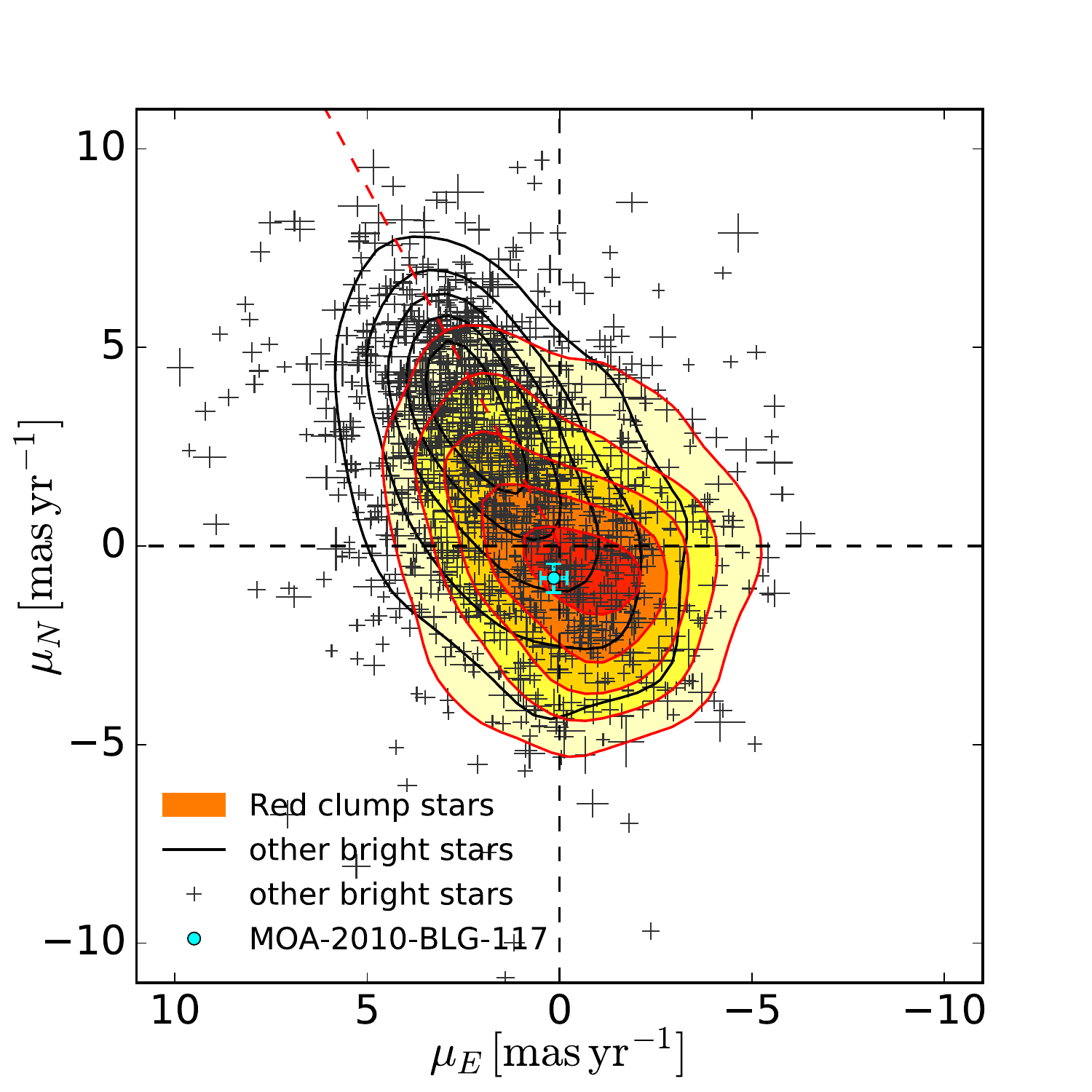}
\caption{The proper motions of stars brighter than the source stars plus blend from the 
OGLE-IV survey. The red and yellow shaded contours indicate the proper motion distribution of 922
bulge red clump stars and the black contours curves
indicate the distribution of the remaining 2239 stars brighter
than $I < 17.87$. The red clump stars come from a compact elliptical region of the CMD, elongated in the 
direction of the extinction vector. The dashed red line extending in
the NNE direction indicates the direction of Galactic rotation, so we expect the distribution of Galactic
disk stars to be extended in the direction of this line. The cyan colored spot with error bars 
in both the E and N directions indicates the target. The target
which consists of the two bound source stars, and a blend star similar in brightness to each of the two
source stars. The proper motion of the source stars indicates that they are likely to be
bulge stars.
\label{fig-ogle_pm}}
\end{figure}

Another indication that the source stars must reside in the Galactic bulge comes from
the proper motion of the source star system.
\citet{skowron14} has developed a method to determine the proper motion of 
microlens source stars in the presence of a modest amount of blending with
other stars. We have used this method to measured the proper motions of stars
brighter than $I < 17.87$, for just over 5 years of OGLE-IV data. (This magnitude cut
is two magnitudes below the red clump centroid.) Figure~\ref{fig-ogle_pm} shows 
that the proper motion of the target, consisting of the two source stars and a blend
stars with a magnitude of about the average of the two source stars. If the blend
star was the lens, it would be in the Galactic disk, so we would expect that the
average proper motion of the two source stars and the blend to be shifted slightly
in the direction of the disk rotation (given by the dashed white line in the NNE direction).
Instead, we find that the proper motion of the target to be 
$(\mu_{\rm E,E}, \mu_{\rm E,N}) = (0.15\pm 0.34, -0.81\pm 0.36)\,$mas/yr.
This clearly indicates that the target is unlikely to be in the disk. Of course, it could
be that the blend star and the two source stars are not in the same population, and their
proper motions could partially cancel. However, out light curve modeling indicates
that the lens-source relative proper motion is in the range 2-$3\,$mas/yr, so if the
blend star was the lens, its proper motion could be at most $\sim 1\,$mas/yr in the
direction of disk rotation. Thus, it would not have disk kinematics. This tends to confirm our
conclusion that the blend star cannot be the lens.

So, we conclude that the source star system resides in the Galactic bulge and that the host
star mass and the lens system distance are determined by the $\pi_E$ and $\theta_E$ 
measurements, as 
described in Section~\ref{sec-lens_prop}. 

\section{Discussion and Conclusions}
\label{sec-conclude}

We have presented the first planetary microlensing event with two magnified source stars.
This event has an obvious planetary feature, but it could not be modeled with a single
source star microlensed by a lens system consisting of one star and one planet. The basic
properties of the planetary feature could be explained by models with two source stars or else a 
circumbinary planet. The choice between these two options was delayed by early difficulties
in modeling the event. These difficulties were overcome by adding the requirement 
that the flux ratio between the two source be consistent
with different passbands allows the best light curve model to be found much more easily.
The finite source effects and microlensing parallax signal indicate that the planet and
host have masses of $m_p = 0.58\pm 0.10 M_{\rm Jup}$ and $M_h =0.51 \pm 0.07 \msun$
at a two-dimensional separation of $a_\perp = 2.44\pm 0.26\,$AU and a distance of
$D_L  = 3.4\pm 0.2\,$kpc. This is a Jupiter mass-ratio planet orbiting at about twice
the distance of the snow line, which is similar to Jupiter's orbit.

One complication in the interpretation of this event is the $K$-band Keck AO images
that indicate an excess of flux at the location of the source. This excess flux is much
brighter than the brightness expected from the lens star, based on the mass determined
from the $\theta_E$ and $\pi_E$ measurements. We consider possibility that this excess
flux could be due to the lens, but we find that the excess flux is more likely to be due
to a companion to the lens star, the source stars, or an unrelated star. This is not the
first planetary microlensing event with a binary source star, as the planetary event
OGLE-2007-BLG-368 \citep{sumi10} has a binary source star that was revealed via
the xallarap effect. (Xallarap is the effect of source orbital motion on the microlensing
light curve.)

This event was as challenging to model as events with an additional lens mass,
either a second star \citep{gould14,poleski14,bennett16}
or a second planet \citep{gaudi-ogle109,bennett-ogle109,han_ob120026,beaulieu16}.
However, events with an additional lens mass
have interesting implications regarding the properties of exoplanet systems, while
events with two source stars do not. The only advantage of a second source star is
a modest increase in the exoplanet detection efficiency. Nevertheless, microlensing is
currently our best method for understanding the population of exoplanets that
orbit beyond the snow line, and the statistical analysis of the
planet populations probed by the microlensing method requires the correct
microlensing model be found for all planetary microlensing events. The
new method that we have presented in this paper aids in this effort, and it
has enabled the MOA Collaboration analysis that has discovered a break
in the exoplanet mass ratio function \citep{suzuki16}.

\acknowledgments 
D.P.B., A.B., and D.S.\  were supported by NASA through grant NASA-NNX12AF54G.
This work was partially supported by a NASA Keck PI Data Award, administered by the NASA
Exoplanet Science Institute. Data presented herein were obtained at the W. M. Keck Observatory
from telescope time allocated to the National Aeronautics and Space Administration through the
agencyÕs scientific partnership with the California Institute of Technology and the University of
California. The Observatory was made possible by the generous financial support of the W. M.
Keck Foundation.
The OGLE Team thanks Profs. M. Kubiak and G. Pietrzy{\'n}ski for their
contribution to the collection of the OGLE photometric data. The OGLE
project has received funding from the National Science Centre, Poland,
grant MAESTRO 2014/14/A/ST9/00121 to A.U.
Work by C.R.\ was supported by an appointment to the NASA Postdoctoral Program at the 
Goddard Space Flight Center, administered by USRA through a contract with NASA.
Work by N.K.\ is supported by JSPS KAKENHI Grant Number JP15J01676.
A.G.\ and B.S.G.\ were supported by NSF grant AST 110347 and
by NASA grant NNX12AB99G.

\end{document}